%% file: main.tex
\documentclass[conference]{IEEEtran}
\usepackage{cite}
\usepackage{amsmath,amssymb,amsfonts}
\usepackage{algorithmic}
\usepackage{graphicx}
\usepackage{textcomp}
\usepackage{xcolor}
\usepackage{verbatim}
\usepackage{subcaption}
\usepackage{fancyhdr}
\usepackage[utf8]{inputenc}

\def\BibTeX{{\rm B\kern-.05em{\sc i\kern-.025em b}\kern-.08em
    T\kern-.1667em\lower.7ex\hbox{E}\kern-.125emX}}

\usepackage{enumitem}

\newcommand{\R}{\mathbb{R}}

\begin{document}

\pagenumbering{gobble}

\title{Cross-Stack Workload Characterization of Deep Recommendation Systems 
}

\author{Samuel Hsia$^1$, Udit Gupta$^{1,2}$, Mark Wilkening$^1$,
\\ Carole-Jean Wu$^2$, Gu-Yeon Wei$^1$, David Brooks$^1$ 
\\ \\ $^1$Harvard University\,\,\,\,\,\,\,\,\,\,$^2$Facebook Inc.
\\ \\ shsia@g.harvard.edu
}

\maketitle



\pagestyle{plain}


\begin{abstract}

Deep learning based recommendation systems form the backbone of most personalized cloud services. Though the computer architecture community has recently started to take notice of deep recommendation inference, the resulting solutions have taken wildly different approaches -- ranging from near memory processing to at-scale optimizations. To better design future hardware systems for deep recommendation inference, we must first systematically examine and characterize the underlying systems-level impact of design decisions across the different levels of the execution stack.
In this paper, we characterize eight industry-representative deep recommendation models at three different levels of the execution stack: algorithms and software, systems platforms, and hardware microarchitectures. Through this cross-stack characterization, we first show that system deployment choices (i.e., CPUs or GPUs, batch size granularity) can give us up to $15\times$ speedup. To better understand the bottlenecks for further optimization, we look at both software operator usage breakdown and CPU frontend and backend microarchitectural inefficiencies. Finally, we model the correlation between key algorithmic model architecture features and hardware bottlenecks, revealing the absence of a single dominant algorithmic component behind each hardware bottleneck.

\end{abstract}

\input{text/intro_ug}
\input{text/background}
\input{text/methodology}
\input{text/speedup}
\input{text/software}
\input{text/hardware}
\input{text/discussion}

\input{text/related_work}
\input{text/conclusion}
\input{text/acknowledgements}

\bibliographystyle{IEEEtran}
\bibliography{references}

\end{document}

%% file: text/intro_ug.tex
\section{Introduction}~\label{sec:introduction}
\begin{figure}[t]
    \centering
    \includegraphics[width=\linewidth]{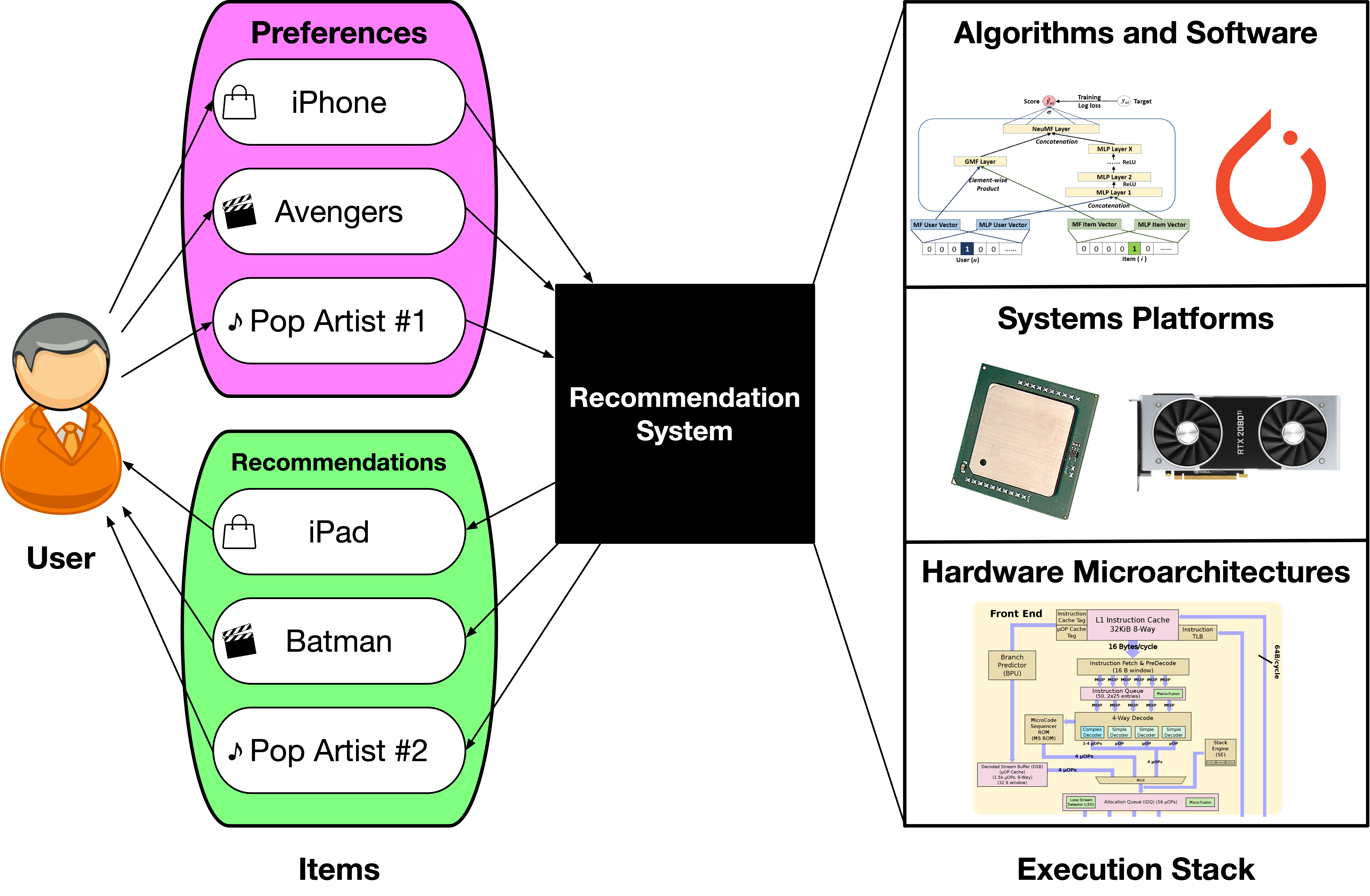}%
    \caption{\textit{Recommendation systems} leverage information about a user's preferences to recommend new content. The three main components include users, items, and the recommendation system itself.
    This paper analyzes recommendation inference at three different levels of the computing stack: algorithms and software, systems platforms, and microarchitectures. 
    }
    \label{fig:rec_overview}
    \vspace{-1em}
\end{figure}

Recommendation systems recommend items to users based on the users' personal preferences.
E-commerce marketplaces (e.g., Amazon, Alibaba) that recommend relevant goods for purchase~\cite{linden2003amazon, zhou2018din, zhou2019dien}, 
social media platforms (e.g., Facebook, Twitter) that regularly update user feeds with new multimedia content\cite{naumov2019dlrm}, 
and entertainment services (e.g., Netflix, YouTube) that promote new playlists~\cite{he2017ncf, zhao2019mtwnd} all heavily rely on high-accuracy recommendation systems to maintain quality of service and positive user experience. 
To achieve the highest possible recommendation accuracies, these recommendation algorithms have evolved from using classical information filtering techniques \cite{Sarwar2001CF, funk2006matrix} to state-of-the-art deep-learning based solutions. Figure \ref{fig:rec_overview} depicts the general workflow of a typical recommendation system in the context of the aforementioned internet-based applications -- these algorithms are programmed using different deep learning frameworks and executed on systems of different architectures.

While deep-learning solutions provide high quality recommendations, they also require high infrastructure overheads to run efficiently.
For instance, Facebook's recommendation use cases require more than 10$\times$ the datacenter inference capacity compared to computer vision and natural language processing tasks~\cite{kim2018hpca}.
As a result, over 80\% of machine learning inference cycles on Facebook's datacenter fleet are devoted to recommendation filtering and ranking~\cite{gupta2020architectural}. The model training process tells a similar story --- over 50\% of the training demands are attributed to deep recommendation models\cite{naumov2020fbtraining}.
Similar capacity demands can be found at other companies like Google~\cite{jouppi2017datacenter}, Amazon~\cite{mckinsey2018rec},  Alibaba~\cite{zhou2018din, zhou2019dien}, and Baidu~\cite{zhao_2019}. 

Given the importance of recommendation models, exploring custom hardware solutions will be important for further workload acceleration and infrastructure efficiency.
Recent work has demonstrated that some recommendation models -- given their large memory capacity requirements and irregular memory access patterns -- pose unique performance bottlenecks compared to other deep learning based workloads (e.g., CNNs and RNNs)~\cite{gupta2020deeprecsys}.
Thus, existing proposals for accelerating datacenter scale CNNs and RNNs do not directly apply to recommendation models.
By exploiting these unique compute characteristics, researchers have demonstrated that novel memory systems can provide significant speedup on a specific class of recommendation models dominated by table lookup operations~\cite{rhu2019tensordimm, ke2019recnmp, hwang2020centaur}.
However, there still exists large algorithmic diversity across recommendation use cases \cite{zhang2019recsurvey, gupta2020deeprecsys, mlperf-reco-advisory}.
Because of this diversity, holistic architectural bottleneck analysis -- for both CPU microarchitecture and heterogeneous hardware -- will enable hardware customization for improving recommendation inference efficiency.

In this paper, we perform a detailed workload characterization of recommendation inference at three different levels of the execution stack: algorithms and software, systems platforms, and hardware microarchitectures.
First, we evaluate the eight recommendation networks on a variety of server class CPUs and AI accelerators (i.e., GPUs).
This evaluation shows the optimal system -- between CPUs and GPUs and between microarchitecture generations (i.e., Intel Broadwell versus Cascade Lake and NVIDIA Pascal versus Turing) -- varies based on the model architectures and inference batch sizes. 
In addition to heterogeneous system evaluation, we demonstrate that Caffe2 software operator usage breakdowns -- which vary across model architectures, batch sizes, and underlying hardware -- expose performance bottlenecks and opportunities for architectural optimizations.
Finally, we perform a detailed microarchitectural performance analysis on server class Intel Broadwell and Cascade Lake CPUs using TopDown~\cite{yasin2014topdown}.
Our microarchitectural analysis reveals that recommendation inference suffers from a variety of microarchitecture inefficiencies related to frontend decoders, backend functional units and memory systems, and branch speculation.

%

The main contributions of this paper are:

\begin{enumerate}
  

  \item 
  While conventional wisdom and recent research indicates that AI accelerators, such as GPUs, readily accelerate deep learning workloads, this work shows the optimal hardware system (i.e., CPU versus GPU) varies based on recommendation use cases (Section \ref{sec:speedup}).
  \item In addition to overall execution times, we analyze software operator usage at the algorithm level. Previous work classifies recommendation models based on their operator breakdowns at fixed use cases on CPUs. In contrast, our analysis shows that varying the model architecture, batch-size, and hardware platform can alter the target operators for future hardware optimization
  (Section \ref{sec:software}).
  

  
  \item 
  Based on the detailed TopDown results from a server class Intel Broadwell CPU, we see that pipeline bottlenecks vary based on model architecture features.
  FC-intensive models (i.e., RM3, WnD, MT-WnD) suffer from insufficient functional units; embedding-intensive models (i.e., RM1, RM2) suffer from ineffective frontend decoders; 
  attention-based models (i.e., DIN, DIEN) suffer from instruction cache misses (Section~\ref{sec:hardware}).
  
  \item We compare the microarchitectural performance characteristics of Broadwell and Cascade Lake CPUs. 
  The Cascade Lake microarchitecture improves recommendation inference performance across all use cases with its wider Single Instruction Multiple Data (SIMD) execution units and enhanced speculation capabilities.
  With these optimizations, the performance bottleneck on Cascade Lake shifts to the backend memory subsystem (Section~\ref{sec:hardware}).


\end{enumerate}







%% file: text/background.tex
\section{Recommendation Background}~\label{sec:background}
Recommendation is the task of suggesting new content to users based on their preferences and prior content interactions. Recommendation is used in many popular internet services (e.g., search, entertainment streaming, e-commerce) to improve user experience by enabling personalization. 
The main challenge for recommendation systems is accurately modeling users' preferences based on sparse training data --- users often explore only a tiny fraction of all available items. 
This section explores how recommendation systems have evolved over time. 
More specifically, we discuss the semantics of classical recommendation approaches and important algorithmic components of state-of-the-art deep learning methods.


\begin{figure}[t]
    \centering
    \includegraphics[width=\linewidth]{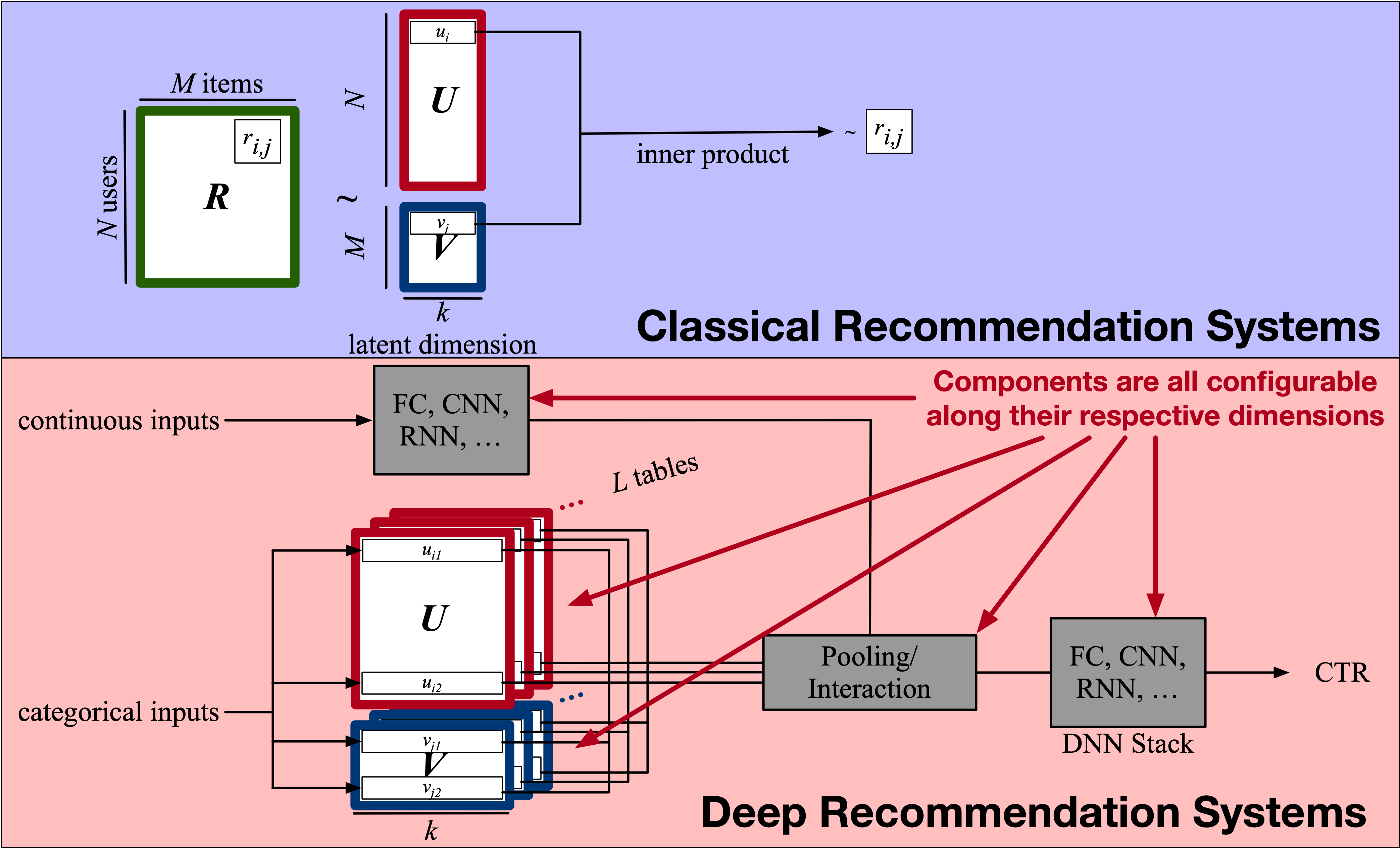}
    \caption{\textbf{Evolution of Recommendation Systems.} Traditional collaborative filtering (\textbf{top}) models user preferences as two (user and item) embedding tables and inner product between table entries. Deep learning based methods (\textbf{bottom}) still leverage embedding tables -- though the number of tables, number of lookups per table, latent dimension, and subsequent matrix operations (gray blocks) are all highly configurable.}
    \label{fig:model_evolution}
    \vspace{-1em}
\end{figure}

\input{tables/rec_models_wide}

\subsection{Classical Recommendation Systems}
Classical recommendation systems are generally categorized as either content-based filtering or collaborative filtering.
Content-based filtering recommends content based on a user's personal preferences and item interaction history while collaborative filtering exploits preference similarity across users.
Content-based filtering models each user as feature vector $\mathbf{u}_i \in \R^{k}$ and each item as feature vector $\mathbf{v}_j \in \R^{k}$, where $k$ is the number of learned features (i.e., latent dimension) for each user and item. 

Collaborative filtering represents historical interactions between $N$ users and $M$ items as a partially-observed matrix $R \in \R^{N \times M}$, where $r_{ij}$ is the historical interaction between user $i$ and item $j$.
To compensate for unobserved entries, collaborative filtering approximates $R$ with matrix factorization (MF) as a user matrix $U \in \R^{N \times k}$ and item matrix $V \in \R^{M \times k}$, where $k$ is again the latent dimension: 
\begin{align}
    R \approx \hat{R} \equiv UV^T
    \label{eqn:mf}
\end{align}
As shown in Figure \ref{fig:model_evolution} (top), to predict the interaction score of user $i$ with item $j$ ($r_{ij}$) we would have to find the inner product of the $i^{th}$ row of $U$ and the $j^{th}$ row of $V$: 
\begin{align}
    r_{ij} \approx \hat{r_{ij}} \equiv {\mathbf{u}_i}^T{\mathbf{v}_j}
    \label{eqn:mf_inference}
\end{align}


\subsection{Deep Recommendation Systems}
In order to leverage the abundance of user data and model complex user-item interactions, recommendation systems have shifted from the aforementioned techniques, e.g., \cite{Sarwar2001CF, funk2006matrix}, to deep-learning based approaches\cite{zhang2019recsurvey}. 
Figure~\ref{fig:model_evolution} (bottom) shows the general deep-learning based model architecture; the embedding tables and components shaded in gray (i.e., DNN-stacks and pooling/interaction layers) are highly configurable:

\begin{itemize}[leftmargin=*]
    \item \textbf{Embedding Tables.} Deep recommendation systems rely on embedding tables (Figure~\ref{fig:model_evolution}, red and blue outline) to encode information about users and items. Every input sample has categorical components (e.g., movies a user likes) -- represented as multi-hot encoded vectors -- that are processed as table lookups. 
    Embedding tables are configured based on the
    number of lookups, number of entries per table, number of tables, and latent dimension of embedding vectors. 
    As table sizes increase ($\sim$GBs), access patterns become increasingly sparse -- leading to irregular memory accesses that make system optimizations challenging.
    \item \textbf{DNN Stacks.} In addition to categorical features, input samples also include continuous features (e.g., user age) that are processed directly by DNN stacks. DNN stacks range from vanilla fully-connected (FC) layers to more complicated architectures (e.g., autoencoders, CNNs, RNNs).
    \item \textbf{Feature Interaction Layers.} To unify outputs from embedding table lookups and DNN stacks, deep recommendation systems have feature interaction layers that range from simple concatenation to attention mechanisms.
\end{itemize}

%% file: tables/rec_models_wide.tex
\begin{table*}[t]
\begin{center}
\resizebox{0.9\linewidth}{!}{\begin{tabular}{|c|c|c|c|}
\hline
\textbf{Model} & \textbf{Application Domain (Evaluation)}                                       & \textbf{Unique Requirement/Use Case}                                                                                            & \textbf{Model Architecture Insight}                                                                                                                           \\ \hline
NCF            & \begin{tabular}[c]{@{}c@{}}Movies \\ (MovieLens)\end{tabular}                  & \begin{tabular}[c]{@{}c@{}}Small amount of required training data\\ (see \# of embedding tables)\end{tabular}                   & Small model with only four embedding tables                                                                                                                   \\ \hline
RM1            & \begin{tabular}[c]{@{}c@{}}Social Media \\ (Facebook)\end{tabular}             & Early stage filtering (i.e., low run-time requirements)                                                                         & \begin{tabular}[c]{@{}c@{}}Small model with medium amount (80) \\ of lookups per embedding table\end{tabular}                                                 \\ \hline
RM2            & \begin{tabular}[c]{@{}c@{}}Social Media \\ (Facebook)\end{tabular}             & \begin{tabular}[c]{@{}c@{}}Late stage ranking (i.e., high accuracy requirements) \\ targeting categorical features\end{tabular} & \begin{tabular}[c]{@{}c@{}}Large model with large amount (120) \\ of lookups per embedding table\end{tabular}                                                 \\ \hline
RM3            & \begin{tabular}[c]{@{}c@{}}Social Media \\ (Facebook)\end{tabular}             & \begin{tabular}[c]{@{}c@{}}Late stage ranking (i.e., high accuracy requirements) \\ targeting continuous features\end{tabular}  & \begin{tabular}[c]{@{}c@{}}Large model with large FC stacks \\ and immediate continuous input processing\end{tabular}                                         \\ \hline
WnD            & \begin{tabular}[c]{@{}c@{}}Smartphone Applications \\ (Google Play Store)\end{tabular} & \begin{tabular}[c]{@{}c@{}}Generic large-scale regression and classification problems\\  with categorical features\end{tabular} & Medium model with large FC stacks                                                                                                                             \\ \hline
MT-WnD         & \begin{tabular}[c]{@{}c@{}}Video \\ (YouTube)\end{tabular}                     & Evaluation of multiple objectives (e.g., likes, ratings)                                                                        & \begin{tabular}[c]{@{}c@{}}Large model with multiple parallel FC stacks\\  on top of WnD\end{tabular}                                                         \\ \hline
DIN            & \begin{tabular}[c]{@{}c@{}}E-Commerce \\ (Alibaba)\end{tabular}                & \begin{tabular}[c]{@{}c@{}}Model evolving user preferences \\ (i.e., time-series nature of dataset)\end{tabular}                & \begin{tabular}[c]{@{}c@{}}Large model with local activation weights\\  for large amount (750) of lookups \\ from user behavior embedding tables\end{tabular} \\ \hline
DIEN           & \begin{tabular}[c]{@{}c@{}}E-Commerce \\ (Alibaba - Taobao)\end{tabular}       & \begin{tabular}[c]{@{}c@{}}Model evolving user preferences\\ (i.e., time-series nature of dataset)\end{tabular}                 & \begin{tabular}[c]{@{}c@{}}Medium model with interaction GRUs \\ to replace large amount of lookups found in DIN\end{tabular}                                 \\ \hline
\end{tabular}}
\end{center}
\caption{Summary of eight industry-representative recommendation models and their important architectural insights.}
\label{tab:rec_models}
\vspace{-1em}
\end{table*}

%% file: text/methodology.tex
\section{Methodology}~\label{sec:methodology}

Personalized recommendation systems run a diverse collection of state-of-the-art deep-learning models across heterogeneous datacenter hardware.
To understand the impact of algorithmic model diversity on inference performance, we characterize eight industry-representative, publicly-available recommendation models.
The model implementations are from the open-sourced DeepRecSys repository and are also not pre-trained as this study focuses solely on inference compute requirements~\cite{gupta2020deeprecsys}.
We characterize the recommendation model performance on server class CPUs (i.e., Intel Broadwell and Cascade Lake) and GPU-based AI accelerators (i.e., NVIDIA 1080 Ti and T4).
This section describes the models and system platforms used in this work.

\subsection{Deep Recommendation Models}

Figure \ref{fig:model_evolution} (bottom) provides a generalization of deep recommendation model architectures.
Building on the general model architecture, our characterization studies eight industry-representative deep recommendation models with unique network parameters, as shown in Table~\ref{tab:rec_models}~\cite{gupta2020deeprecsys}.

\input{tables/hw}

\begin{enumerate}[leftmargin=*]

\begin{figure*}[ht!]
    \centering
    \includegraphics[width=0.95\linewidth]{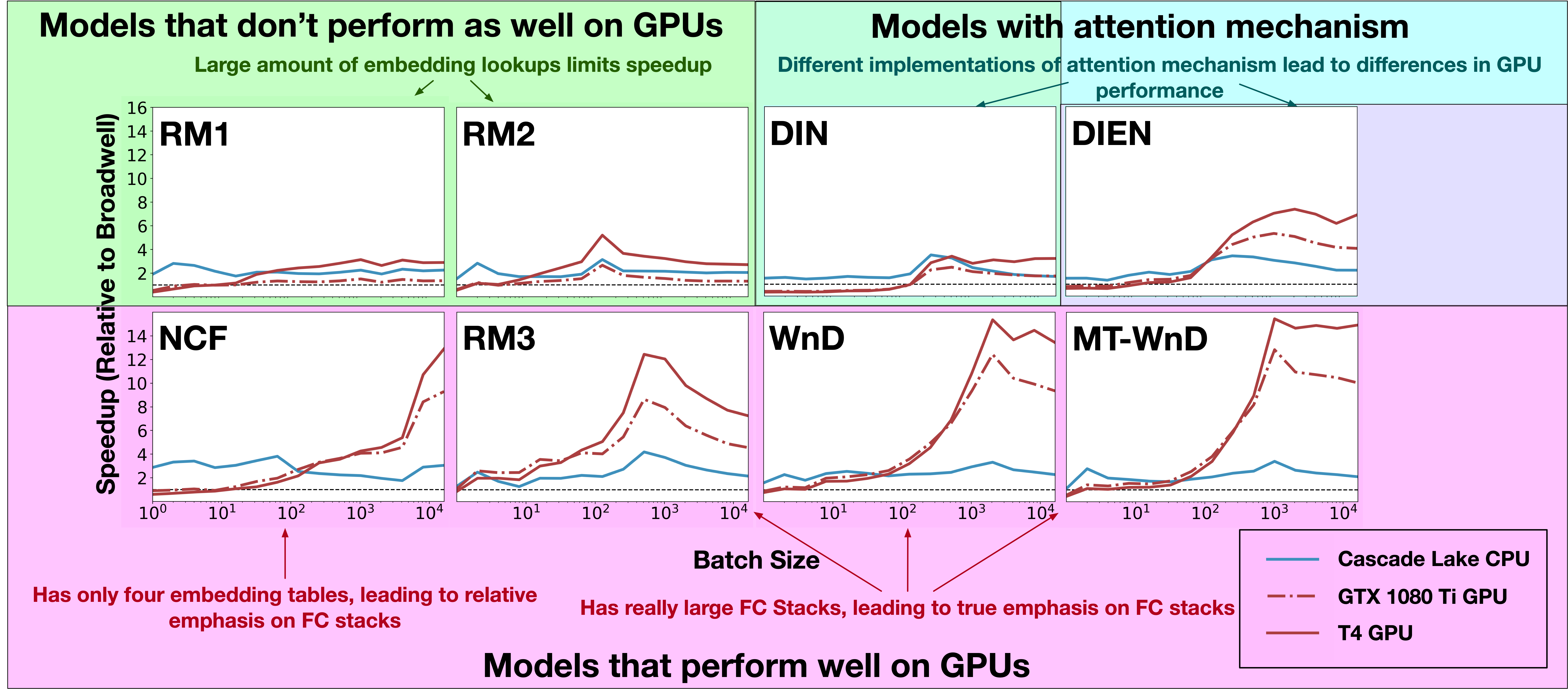}
    \caption{\textbf{Systems performance evaluation} represented as speedup over Broadwell CPU across models, batch-sizes, and hardware platforms. Models are grouped into three primary, overlapping categories -- ones that perform well on GPUs, ones that have comparable performance on CPUs and GPUs, and attention-based models with varying implementations.}
    \label{fig:speedups}
    \vspace{-1em}
\end{figure*}

\item \textbf{Neural Collaborative Filtering (NCF)} extends matrix factorization with multi-layer perceptrons (MLPs) and non-linearities: 
\begin{equation}
    r_{ij} \approx \hat{r_{ij}} \equiv \phi(\mathbf{w}^T({\mathbf{u}_i}\circ{\mathbf{v}_j}))
    \label{eqn:gmf}
\end{equation}
where $\phi$ and $\mathbf{w}$ are the activation function and weights respectively. Although NCF has only four embedding tables, it has shown success with the MovieLens dataset~\cite{he2017ncf}.

\item \textbf{Deep Learning Recommendation Model (DLRM RM1, RM2, RM3)} is a highly configurable model with multi-hot encoded embedding lookups.
Outputs of embedding lookups are aggregated with the output of DNN stacks that process continuous input features.
We configure three representative DLRM networks -- RM1, RM2, RM3 -- with varying ratios of FC weights and embedding lookups based on Facebook's social media ranking models~\cite{naumov2019dlrm, gupta2019architectural, gupta2020deeprecsys}. 

\item \textbf{Wide and Deep (WnD)} captures both the memorization and generalization benefits by concatenating outputs of one-hot encoded embedding lookups with continuous inputs.
The resulting features are then processed with deep feed-forward networks.
WnD has been used to rank applications in Google's Play Store~\cite{cheng2016wnd}.

\item \textbf{Multi-Task Wide and Deep (MT-WnD)} expands upon WnD by adding parallel output FC layers on top of WnD to evaluate multiple objectives. While the other models predict a single engagement objective such as click-through rate (CTR), MT-WnD evaluates multiple objectives such as likes and ratings. 
MT-WnD provides high quality next video recommendations on YouTube \cite{zhao2019mtwnd}.

\item \textbf{Deep Interest Network (DIN)} addresses evolving user preferences by implementing the attention mechanism with local activation units for embedding table lookups. 
User embedding tables process a small number of lookups while item embedding tables process hundreds of lookups. 
DIN has been deployed to great success by Alibaba in its online marketplace for display advertising \cite{zhou2018din}.

\item \textbf{Deep Interest Evolution Network (DIEN)} also addresses evolving user preferences but uses multi-layered gated recurrent units (GRUs) to explicitly separate user preferences from user interaction history. For item embedding tables, this leads to fewer lookups per table as more of the information processing is offloaded to the GRU layers. Like DIN, DIEN has been deployed successfully by Alibaba on its display advertising services (specifically on Taobao) \cite{zhou2019dien}.
\end{enumerate}

\subsection{Systems Platforms}

State-of-the-art recommendation models are deployed across heterogeneous hardware systems in datacenters. 
In fact, exploiting hardware heterogeneity to schedule inferences on optimum platforms based on use cases (i.e., model architecture, inference batch-size) significantly improves recommendation performance~\cite{gupta2020deeprecsys}.
This work mimics this hardware heterogeneity by providing in-depth characterizations on two server class CPUs (i.e., Intel Broadwell and Cascade Lake) and two GPUs (i.e., NVIDIA GTX 1080 Ti and T4).
Table~\ref{tab:hw} summarizes the key architectural features of the platforms.
GPUs are connected to CPUs via PCIe 3.0. All results assume single-threaded inference in Caffe2, and include both data loading and model computation times to capture end-to-end recommendation inference.



%% file: tables/hw.tex
\begin{table}[t]
\begin{center}
\resizebox{\linewidth}{!}{\begin{tabular}{|c||c|c|c|c|}
\hline
\textbf{Machines}                                                               & \textbf{Xeon E5-2697A} & \textbf{Xeon Gold 6242} & \textbf{GTX 1080 Ti} & \textbf{T4} \\ \hline
Microarchitecture                                                               & Broadwell              & Cascade Lake            & Pascal               & Turing      \\ \hline
Frequency                                                                       & 2.6 GHz                & 2.8 GHz                 & 1.48 GHz             & 0.58 GHz    \\ \hline
Cores (SM Count)                                                                & 16                     & 16                      & (28)                 & (40)        \\ \hline
\begin{tabular}[c]{@{}c@{}}SIMD \\ (CUDA Capability)\end{tabular}               & AVX-2                  & AVX-512                 & (6.1)                & (7.5)       \\ \hline
L1 Cache Size                                                                   & 32 KB                  & 32 KB                   & 48 KB                & 64 KB       \\ \hline
L2 Cache Size                                                                   & 256 KB                 & 1 MB                    & 2.75 MB              & 4 MB        \\ \hline
L3 Cache Size                                                                   & 40 MB                  & 22 MB                   & N/A                  & N/A         \\ \hline
\begin{tabular}[c]{@{}c@{}}L2/L3 (L1/L2) \\ Cache Inclusion Policy\end{tabular} & Inclusive              & Exclusive               & (Inclusive)          & (Inclusive) \\ \hline
DRAM Capacity                                                                   & 256 GB                 & 384 GB                  & 11 GB                & 16 GB       \\ \hline
DDR Type                                                                        & DDR4                   & DDR4                    & GDDR5X               & GDDR6       \\ \hline
DDR Frequency                                                                   & 2400 MHz               & 2933 MHz                & 1376 MHz             & 1250 MHz    \\ \hline
DDR Bandwidth                                                                   & 77 GB/s                & 131 GB/s                & 484.4 GB/s           & 320 GB/s    \\ \hline
TDP                                                                             & 145 W                  & 150 W                   & 250 W                & 70 W        \\ \hline
\end{tabular}}
\end{center}
\caption{Summary of hardware platforms studied.}
\label{tab:hw}
\vspace{-1em}
\end{table}

%% file: text/speedup.tex
\section{Systems Platforms Evaluation}~\label{sec:speedup}
\begin{figure*}[t]
    \centering
    \includegraphics[width=0.9\linewidth]{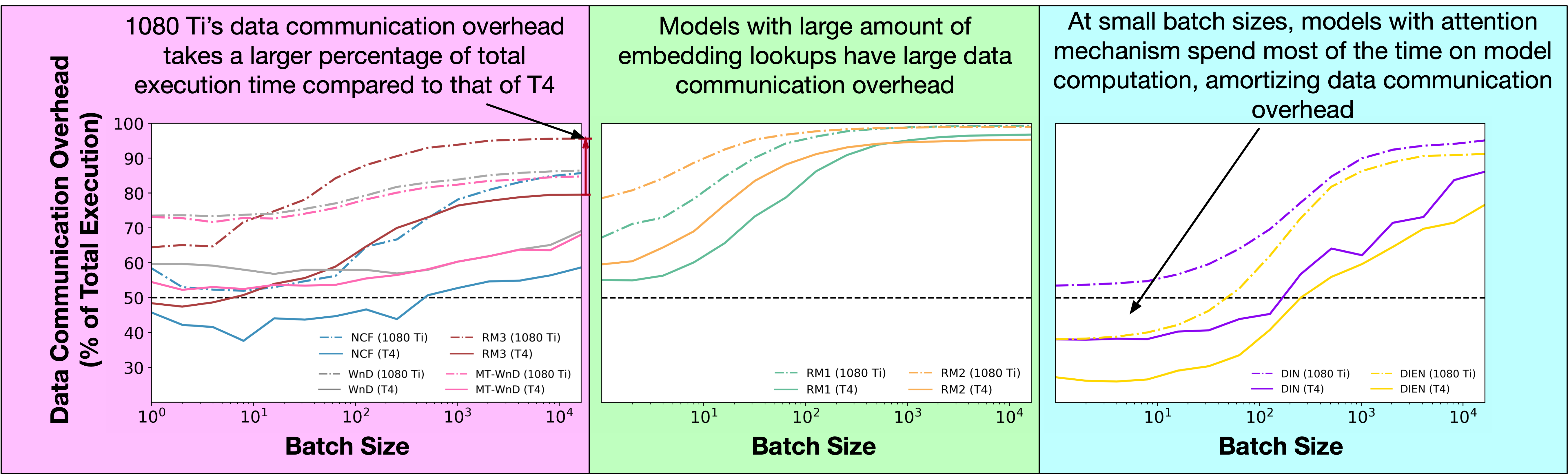}
    \caption{\textbf{GPU data communication overheads} as percent of total execution time. While parts of this overhead can be attributed to software implementation overhead, the majority is due to GPU data loading (i.e., CPU-GPU communication overheads).}
    \label{fig:data_loading}
    \vspace{-1em}
\end{figure*}
This section describes the performance characteristics of the eight recommendation models at different use cases (i.e., input batch sizes and systems platforms). 
The range of batch sizes follows recent work that shows recommendation in datacenters runs with batch sizes from tens to thousands to meet different SLA targets~\cite{gupta2020deeprecsys}; the range of systems platforms exposes the effects of different generations of CPUs and GPUs. 
In the context of GPUs, we find that data-communication overheads limit GPU performance.
Overall, the analysis shows that the optimum hardware for recommendation inference depends on both model architecture and batch size. 



Figure~\ref{fig:speedups} depicts the speedups of the Cascade Lake CPU, GTX 1080 Ti GPU, and T4 GPU over a baseline Broadwell CPU server.
The results are organized by recommendation model, across batch-sizes 1 to 16384, and consider end-to-end execution times (i.e., model-computation plus data-communication). The important observations are: 


\textbf{1) Model architecture plays an important role in accelerating recommendation inference.} 
Models in the bottom row of Figure~\ref{fig:speedups} exhibit high speedup on the NVIDIA 1080 Ti and T4 GPUs.
For larger batch sizes ($\sim10^{3}$), the GPUs provide \textit{an order of magnitude} speedup over the baseline Broadwell CPU; for smaller batch sizes ($<10^{2}$), we observe a $2-4\times$ speedup. 
The relatively high speedups observed come from the models (i.e., NCF, RM3, WND, MT-WND, RM3) sharing an important algorithmic characteristic (Table~\ref{tab:rec_models}): whether it is fewer embedding tables in NCF, large FC stacks for continuous inputs in RM3, or large FC stacks to output final probability scores in WnD and MT-WnD, each of these models relies on FC stacks to model user preferences.
As GPUs readily accelerate matrix operations, they outperform CPUs on NCF, RM3, WND, and MT-WND. 

Compared to the models with large FC components, RM1 and RM2 exhibit relatively lower speedups -- less than 4$\times$ -- when deployed on GPUs (Figure~\ref{fig:speedups} top left).
In fact, at small batch sizes, Cascade Lake consistently outperforms the 1080 Ti GPU (and by at least 2$\times$ at small batch sizes) and offers speedups within 10\% of the T4.
This is a result of RM1 and RM2 having a large number of lookups per embedding table -- 80 and 120 lookups respectively. 
In comparison, the remaining models have fewer than 20 lookups per table.
The larfe number of lookups shifts the performance bottleneck towards embedding operations that comprise of irregular memory accesses (see Section~\ref{sec:software} for details).
Depending on the input batch-size, CPUs and GPUs perform comparably on these models dominated by irregular memory accesses.


\textbf{2) Different model architecture implementations of an algorithmic feature like the attention mechanism can have different hardware implications.} 
Algorithmically, both DIN and DIEN use attention to learn users' evolving interests over time.
DIN implements attention with local activation units and small FC layers followed by concatenation operations for aggregation while DIEN implements attention using multi-layered gated-recurrent units (GRUs).
For DIN, Broadwell machines outperform GPUs at batch-sizes less than 100.
At larger batch-sizes, GPU speedup saturates below 4$\times$.
The lower speedups are a direct result of DIN implementing attention with heavy concatenation operations that perform poorly on GPUs.
In comparison, DIEN achieves up to 7$\times$ speedup on GPUs compared to Broadwell, as GRUs translate to matrix multiplications that perform well on GPUs.

\begin{figure}[t]
    \centering
    \includegraphics[width=0.9\linewidth]{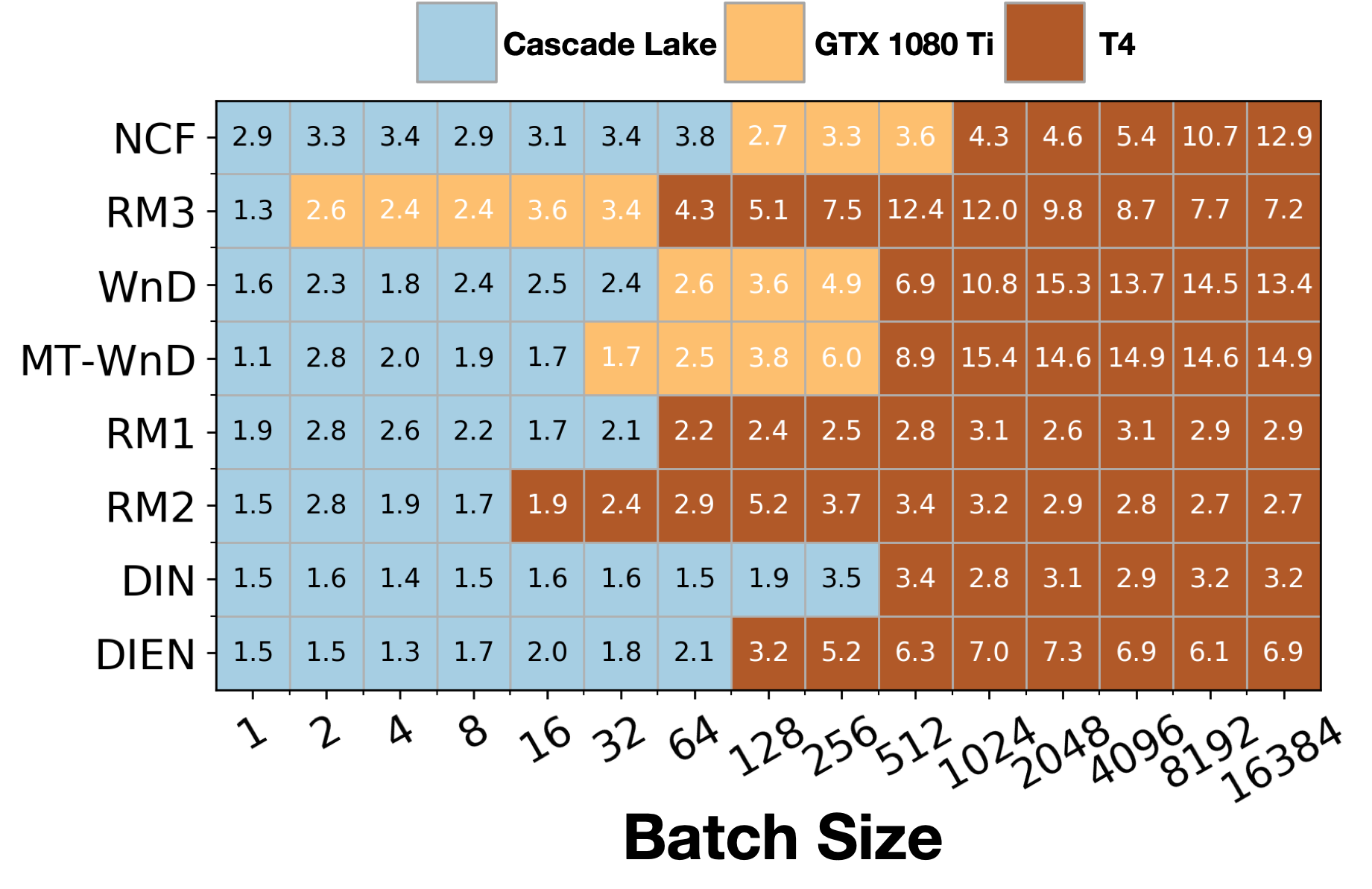}
    \caption{\textbf{Optimal hardware for each model architecture and batch size:} each grid cell show the speedup over Broadwell when using the optimum hardware (color).}
    \label{fig:optimal_hw}
    \vspace{-1em}
\end{figure}


\begin{figure*}[t]
    \centering
    \includegraphics[width=\linewidth]{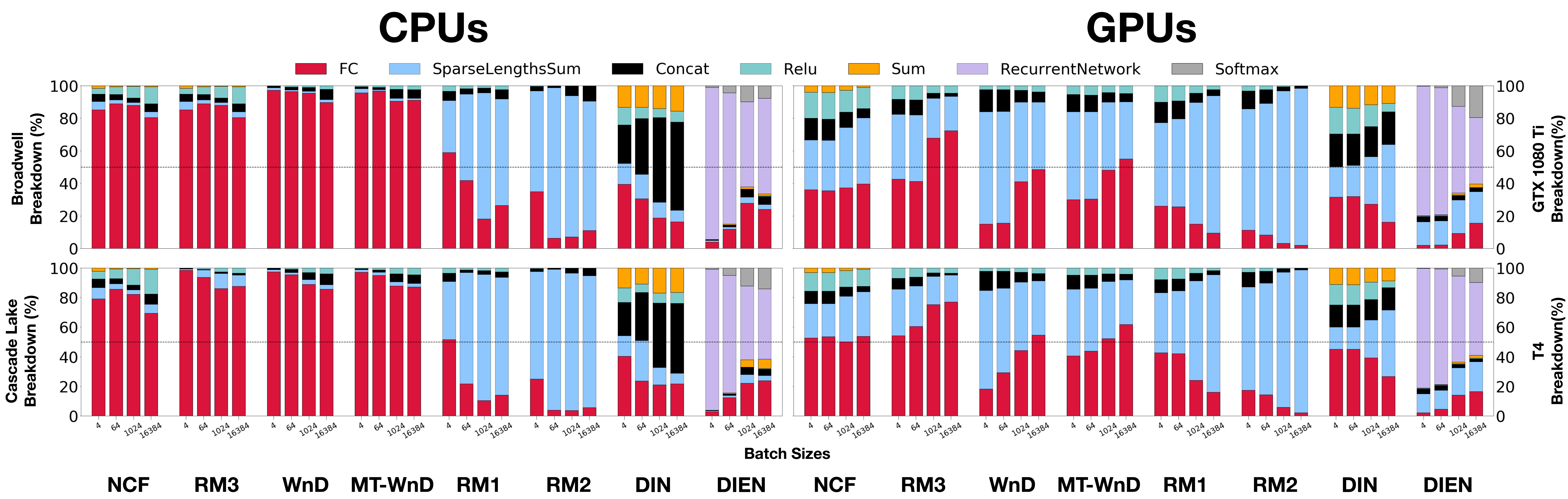}
    \caption{\textbf{Caffe2 operator breakdowns} with CPUs (left) and GPUs (right). Models readily accelerated by GPUs are dominated by matrix operations (i.e., FC in red and recurrent layers in purple).
    Operator breakdowns between CPUs and GPUs vary significantly --- models dominated by FC execution time on CPUs spend a large fraction of time on other operators on GPUs.}
    \label{fig:op_breakdown}
    \vspace{-1em}
\end{figure*}

\textbf{3) Compared to Broadwell, Cascade Lake improves performance across all models and batch sizes.}
Across all use cases, encompassing models and batch-sizes, Cascade Lake achieves higher performance than Broadwell CPUs.
Following Table~\ref{tab:hw}, the improved performance is a result of various micro-architectural features such as wider SIMD width for FC-focused models, larger L2 cache capacity, and higher DRAM frequency.
Section~\ref{sec:hardware} details the micro-architectural features that enable higher performance on Cascade Lake.

\begin{figure}[t]
    \centering
        \includegraphics[width=\linewidth]{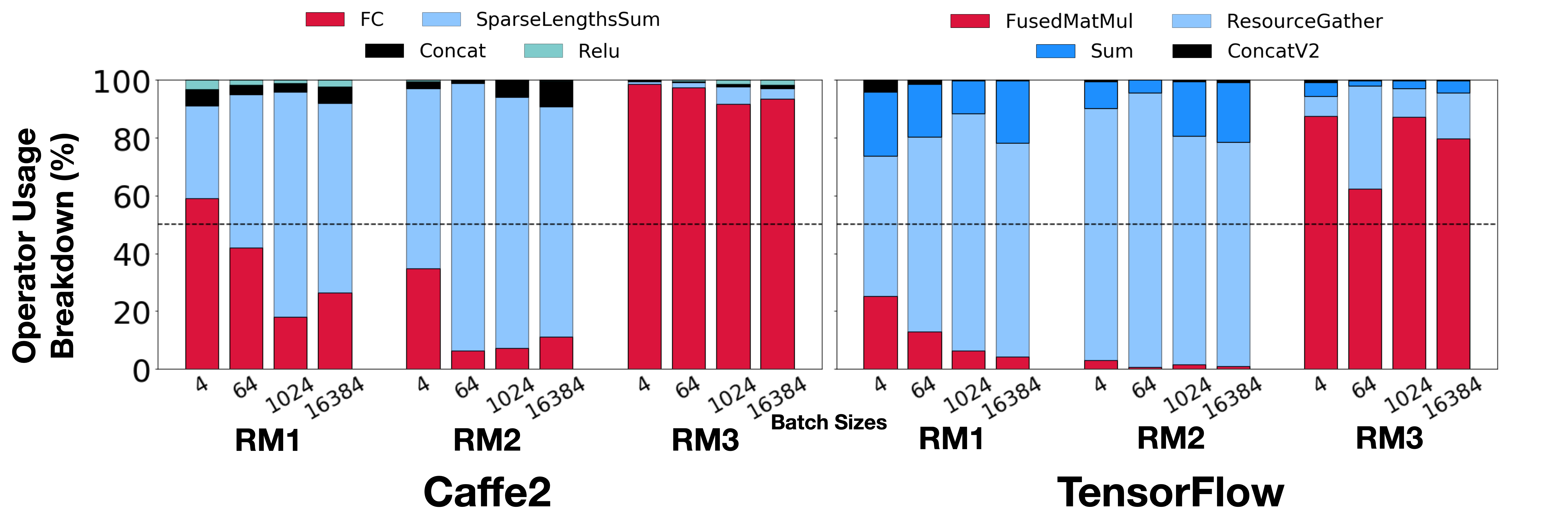}%
    \caption{\textbf{Comparison of Caffe2 and TensorFlow operator breakdowns} for DLRM-based recommendation models. Operators that comprise the majority of execution time are similar across both frameworks. Note, embedding table operations correspond to \texttt{SparseLengthsSum} in Caffe2 and the combination of \texttt{ResourceGather} and \texttt{Sum} in TensorFlow.}
    \label{fig:op_breakdown_tf}
    \vspace{-1em}
\end{figure}

\textbf{4) Compared to GTX 1080 Ti, T4 improves performance for specific models and batch-sizes.}
For NCF, RM3, WnD, MT-WnD, and DIEN, T4 outperforms the 1080 Ti at batch sizes larger than $\sim 10^3$, offering higher speedups due to higher streaming multiprocessor (SM) count. 
However, for RM1 and RM2, T4 becomes advantageous at smaller batch sizes -- due to the increase in GDDR5X to GDDR6 frequency. 
This is important as for latency-critical applications with strict SLA targets, input samples must run at small batch sizes.


\textbf{5) GPU speedup over CPU is limited by data communication overheads.} 
Figure \ref{fig:data_loading} quantifies the fraction of time spent on data communication for different models and batch sizes.
Data communication overheads come from offloading both continuous and categorical inputs via PCIe.
For all models, the fraction of time spent on data communication scales with batch size as compute operations are readily accelerated (sub-linear) but data communication is not.
Exact percentage of time spent on data-communication still depends on the model architecture; models that rely on embedding lookups suffer most.
Given the high data-communication overheads, we conclude that running recommendation models out of the box on GPUs underutilizes the GPUs' compute resources.

Figure \ref{fig:optimal_hw} summarizes the results of Section \ref{sec:speedup} by showing how the optimal system platform (color coded) and speedup (number inside cell) vary across the use cases (models across the rows and batch sizes across the columns).
While this section provides a high level intuition on the tradeoffs between CPUs and GPUs, the following sections dive deeper into the heterogeneity behind Figure \ref{fig:optimal_hw}.

%% file: text/software.tex
\section{Algorithms and Software Characterization}~\label{sec:software}
To better understand the system performance trends, in this section, we provide an algorithmic characterization of the different models and use cases. 
More specifically, we breakdown Caffe2 operators' usage for inference across the eight models and batch sizes. 
Furthermore, we show that the algorithmic bottlenecks are consistent across deep-learning frameworks (i.e., Caffe2 and Tensorflow), demonstrating that the performance trends are fundamental to model architectures.

\subsection{Operator Breakdown}

Operator usage breakdowns allow us to quantify the importance of each operator and compare them across model architectures in a \textit{unified} manner. This is extremely important as model architectures for deep recommendation systems are rapidly evolving. In fact, these model architectures often mirror recent advances in deep learning\cite{zhang2019recsurvey}. Figure \ref{fig:op_breakdown} shows the operator breakdown of the eight models -- implemented in Caffe2 -- across four different batch sizes on Broadwell, Cascade Lake, GTX 1080 Ti, and T4 machines. We see that different generations of hardware (top versus bottom rows) and classes of hardware (left versus right columns) alter operator usage breakdown. The important observations are:


\textbf{1) GPUs accelerate models dominated by \texttt{FC} operators on CPUs but struggle with those bottlenecked by \texttt{SparseLengthsSum} on CPUs.} Since GPUs contain large arrays of SMs that execute matrix multiplication efficiently, models with \texttt{FC}-dominated runtimes show the most acceleration (see: NCF, RM3, WnD, and MT-WnD). 
On the other hand, models with CPU runtimes bottlenecked by the \texttt{SparseLengthsSum} operator do not perform as well on GPUs. 
The \texttt{SparseLengthsSum} operator itself consists of both looking up a specified number of embedding vectors from each table and a subsequent partial sum. 
This becomes an issue for GPUs when the number of lookups per table and the table counts increase, leading to irregular memory access patterns.


\begin{figure}[t]
    \centering
    \includegraphics[width=0.9\linewidth]{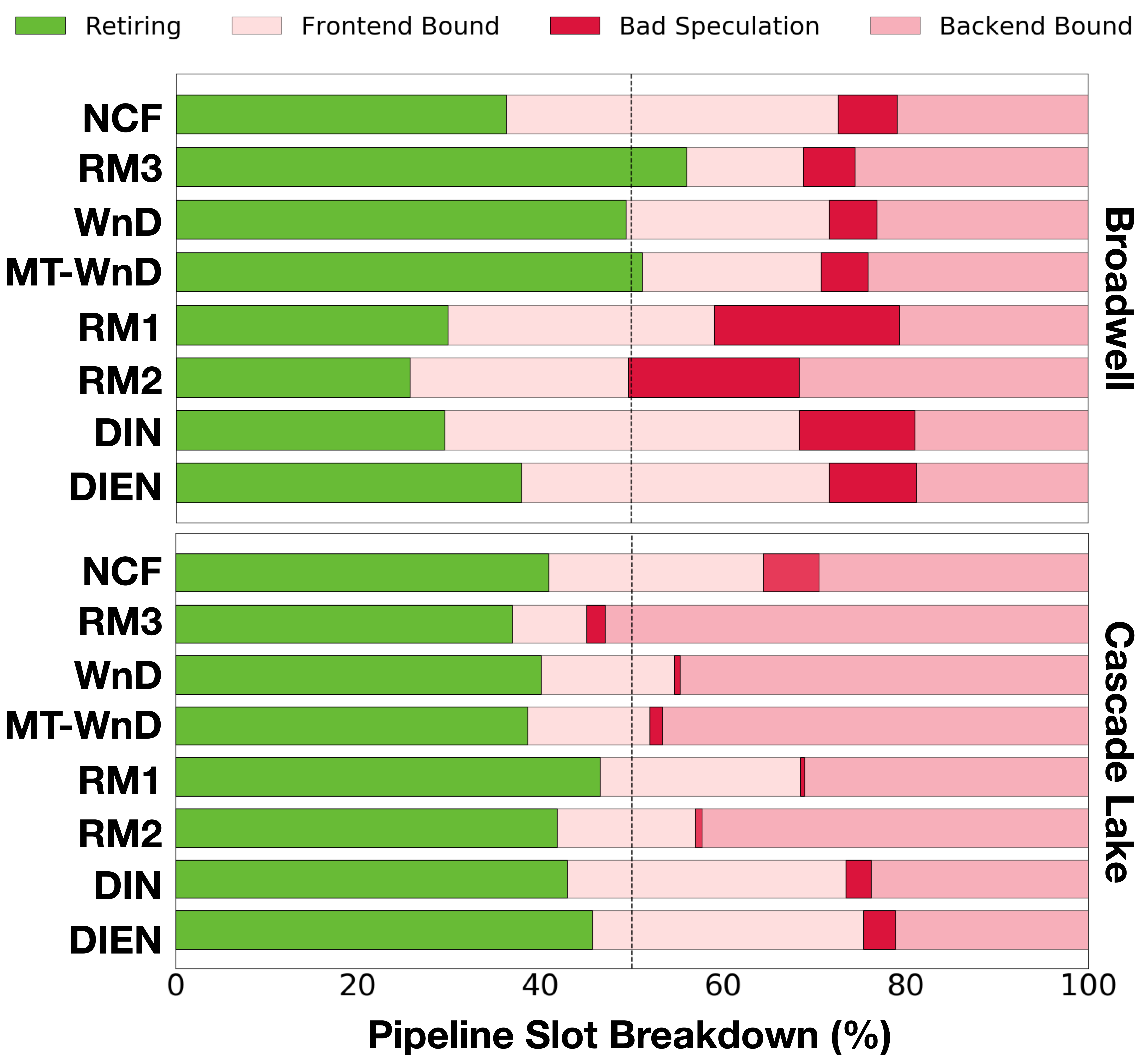}
    \caption{\textbf{TopDown pipeline slot breakdowns.} \textbf{(Top)} On Broadwell, models that rely on matrix operations (i.e., NCF, RM3, WnD, MT-WnD), fill most of their pipeline slots with retiring instructions; the remaining models are either frontend bound or backend bound. 
    \textbf{(Bottom)} On Cascade Lake, models with large FC components (i.e., RM3, WnD, MT-WnD) have fewer retiring  pipeline slots due to wider SIMD width.
    The remaining models exhibit an increase in retiring mainly due to a reduction in bad speculation pipeline slots.}
    \label{fig:td_both}
    \vspace{-1em}
\end{figure}

\textbf{2) In addition to the impact of varying model architecture, batching inference requests across different hardware platforms uncovers additional opportunities for hardware optimization.} Previous work has categorized recommendation models into three types: \textit{MLP-}, \textit{Embedding-}, or \textit{Attention}-dominated models based on using a Broadwell CPU at a fixed batch size of 64 \cite{gupta2020deeprecsys}. While this offers an efficient grouping for high-level discussions, analyzing operator breakdowns across all possible use cases reveals even more optimization points for designing future hardware. For example, on RM1, varying batch sizes from 4 to 64 will shift the dominant operator bottleneck from \texttt{FC} to \texttt{SparseLengthsSum}.
Classifying the recommendation models based on their GPU performance also leads to different conclusions. 
For example, WnD, an FC-heavy model on CPUs, is dominated by the \texttt{SparseLengthsSum} operator at small batch sizes on GPUs.
Identifying these shifting bottlenecks is important in order to thoroughly explore optimization opportunities.
Designing efficient hardware that specializes for low-latency targets (i.e., smaller batch sizes), high-throughput (i.e., larger batch sizes), or other specific cases will require revisiting the operator breakdowns at target use cases.

\textbf{3) Different generations of the same platform type (i.e., CPU/GPU) affect exact operator usages but retain general trends.} On the left subplots of Figure \ref{fig:op_breakdown} are Broadwell and Cascade Lake breakdowns and on the right are GTX 1080 Ti and T4 breakdowns.
Inter-generation microarchitectural changes (i.e. Broadwell to Cascade Lake) affect operator breakdowns (e.g., for RM1 and RM2, time spent on FC layers is reduced) -- Section \ref{sec:hardware} goes more in depth on how microarchitectural differences lead to this.

\subsection{Effects of Different Deep Learning Frameworks}

Figure \ref{fig:op_breakdown_tf} compares operator breakdowns between Caffe2 and TensorFlow for DLRM-based models. 
As the operator breakdowns are similar, we know the optimization targets will be, to first order, the same regardless of differences in software frameworks. The mapping of the operator responsible for FC stacks is straightforward: \texttt{FC} in Caffe2 maps to \texttt{FusedMatMul} in TensorFlow. 
However, the \texttt{SparseLengthsSum} operator in Caffe2 maps to the combination of \texttt{ResourceGather} (lookup) and \texttt{Sum} (pool) operators in TensorFlow.





%% file: text/hardware.tex
\begin{figure}[t]
    \centering
    \includegraphics[width=\linewidth]{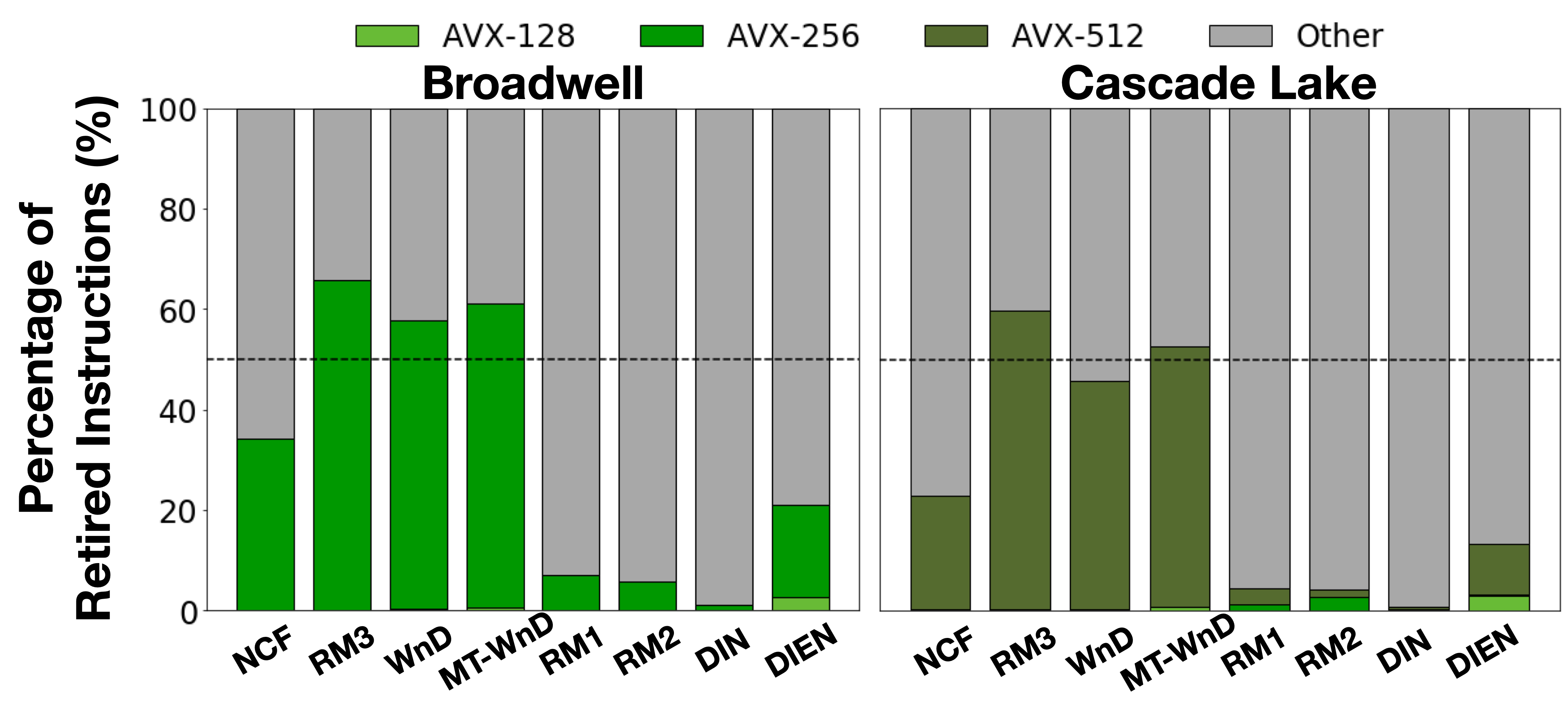}
    \caption{\textbf{Instruction vectorization.} (\textbf{Left}) Broadwell AVX instructions constitute over $60\%$ of retired instructions for models with larger FC layers (i.e., RM3, WnD, MT-WnD). 
    (\textbf{Right}) Cascade Lake's wider SIMD-width results in shorter execution time despite reduced AVX instruction footprint.}
    \label{fig:avx}
    \vspace{-1em}
\end{figure}

\section{CPU Microarchitectural Characterization}~\label{sec:hardware}
Complementing the operator breakdowns, in this section we present a detailed CPU microarchitectural characterization that provides additional insights into the performance trends for recommendation inference. 
In order to better understand the architectural bottlenecks in general purpose processors, we use TopDown-based performance measurement unit (PMU) analysis for server-class Broadwell and Cascade Lake CPUs (Table \ref{tab:hw}) \cite{yasin2014topdown}. 
This analysis shows the important microarchitectural components that form the performance bottlenecks for different recommendation models on Broadwell and how the bottlenecks change for Cascade Lake CPUs.

\subsection{TopDown analysis}
Following TopDown performance analysis~\cite{yasin2014topdown}, we break down the CPU pipeline into four major portions: frontend, speculation, backend, and retiring.
The frontend fetches instructions from memory and converts them into micro-operations ($\mu$ops); speculation realizes predictive optimizations; backend schedules and executes the $\mu$ops; retiring commits the $\mu$ops.
In order to optimize the performance of a processor we must maximize instructions per cycle (IPC).
Generally, IPC can be improved by increasing the fraction of processor cycles devoted to retiring as opposed to stalled in the frontend, speculation, or backend portions. 
Recent work uses TopDown analysis to better understand the fraction of cycles in each pipeline portion for server and datacenter workloads~\cite{kanev2015warehouse, sriraman2019softsku, yasin2014topdown}. 

\subsection{Recommendation performance using TopDown}

Figure \ref{fig:td_both} shows the TopDown breakdown of the eight deep recommendation models, with a batch-size of 16, on Broadwell and Cascade Lake CPUs. 
Generally, on Broadwell, models with larger FC layers (i.e., RM3, WnD, and MT-WND) spend the majority of their cycles in retiring.
On the other hand, the remaining models (i.e., NCF, RM1, RM2, DIN, and DIEN) suffer from a variety of frontend, backend, and bad speculation bottlenecks.
Following are notable observations:

\textbf{1) On Broadwell, larger FC-dominated models benefit from vector execution but remain limited with insufficient functional units.} On Broadwell, models that rely on FC layers (i.e., NCF, RM3, WnD, MT-WND) spend a large percentage of pipeline slots on retiring instructions.
Thus, the natural next step would be to investigate the \textit{degree} of instruction vectorization for these models (Figure \ref{fig:avx}).
On Broadwell, over $60\%$ of all retired instructions for RM3, WnD, and MT-WnD are Advanced Vector Instructions (AVX) (Figure \ref{fig:avx} (Left)). 
This is a result of machine learning frameworks, like Caffe2, translating FC layers to vectorized matrix operations. 

\begin{figure}[t]
    \centering
    \includegraphics[width=\linewidth]{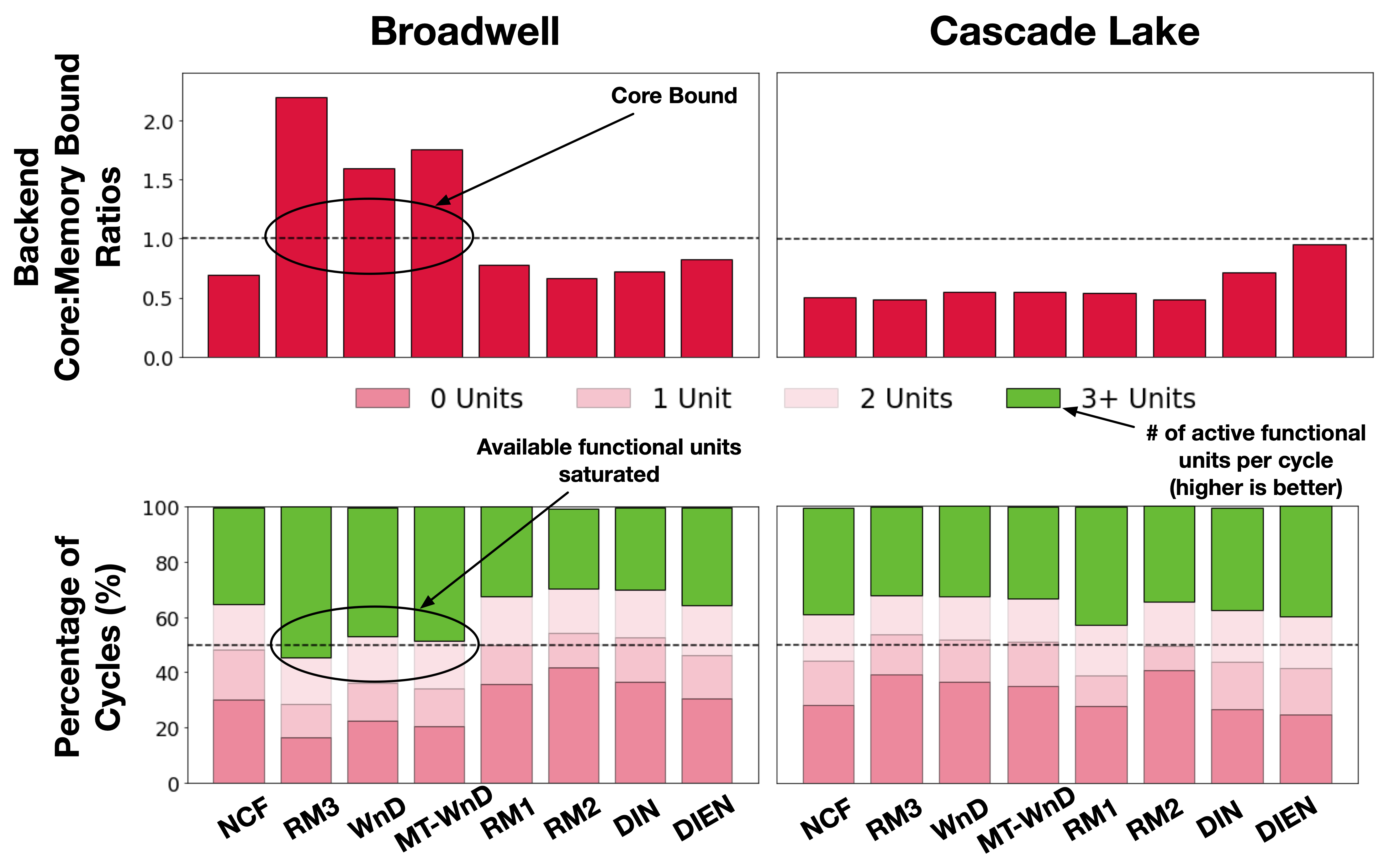}
    \caption{(\textbf{Top}) \textbf{Ratio of Core:Memory Backend Bound cycles}.  Majority of stalls come from functional units on Broadwell and from memory subsystem on Cascade Lake. (\textbf{Bottom}) \textbf{Functional unit usage}. RM3, WnD, and MT-WnD saturate Broadwell's functional units more than other models with a large fraction of cycles that use 3+ units out of 8. Cascade Lake decreases the pressure on functional units.}
    \label{fig:core_bound}
    \vspace{-1em}
\end{figure}

Despite this high degree of vectorization, the larger FC-dominant models still spend a significant fraction of pipeline slots backend bound -- highlighting the need for improved CPU backend pipelines for faster $\mu$ops consumption.
Backend bound cycles can be further classified as either core bound or memory bound. 
Figure \ref{fig:core_bound} (Left) quantifies the core-bound nature of these backend-bound models on a Broadwell machine in two ways. The top row shows the breakdown of the backend bound slots as a core:memory bound ratio, where RM3, WnD, and MT-WnD all show numbers $>1$.
In the case of RM3, where the ratio $\sim2$, there are twice as many functional units-induced stall cycles as memory subsystem-induced stall cycles.
We see that for WnD and MT-WnD, this ratio is $>1.5$.
Thus, despite wide vector execution, larger FC-dominant models remain backend core-bound on Broadwell machines.

Figure \ref{fig:core_bound} (Bottom) details the cycle-level utilization of functional units. 
Recall that Broadwell CPUs have eight functional units: four arithmetic units, two load units, and two store units.
Figure~\ref{fig:core_bound} (left, bottom) shows nearly 50\% of cycles in RM3, WnD, and MT-WnD require more than three functional units: this high functional unit utilization underscores the core-bound bottleneck.
This illustration of the core-bound bottleneck also corroborates the source of GPU speedups for RM3, WnD, and MT-WnD. 
Since these models are bottlenecked by the lack of more functional units, the increase in the amount of compute units (streaming multiprocessors) on GPUs alleviates this core bound issue on Broadwell.

\begin{figure}[t]
    \centering
    \includegraphics[width=0.5\linewidth]{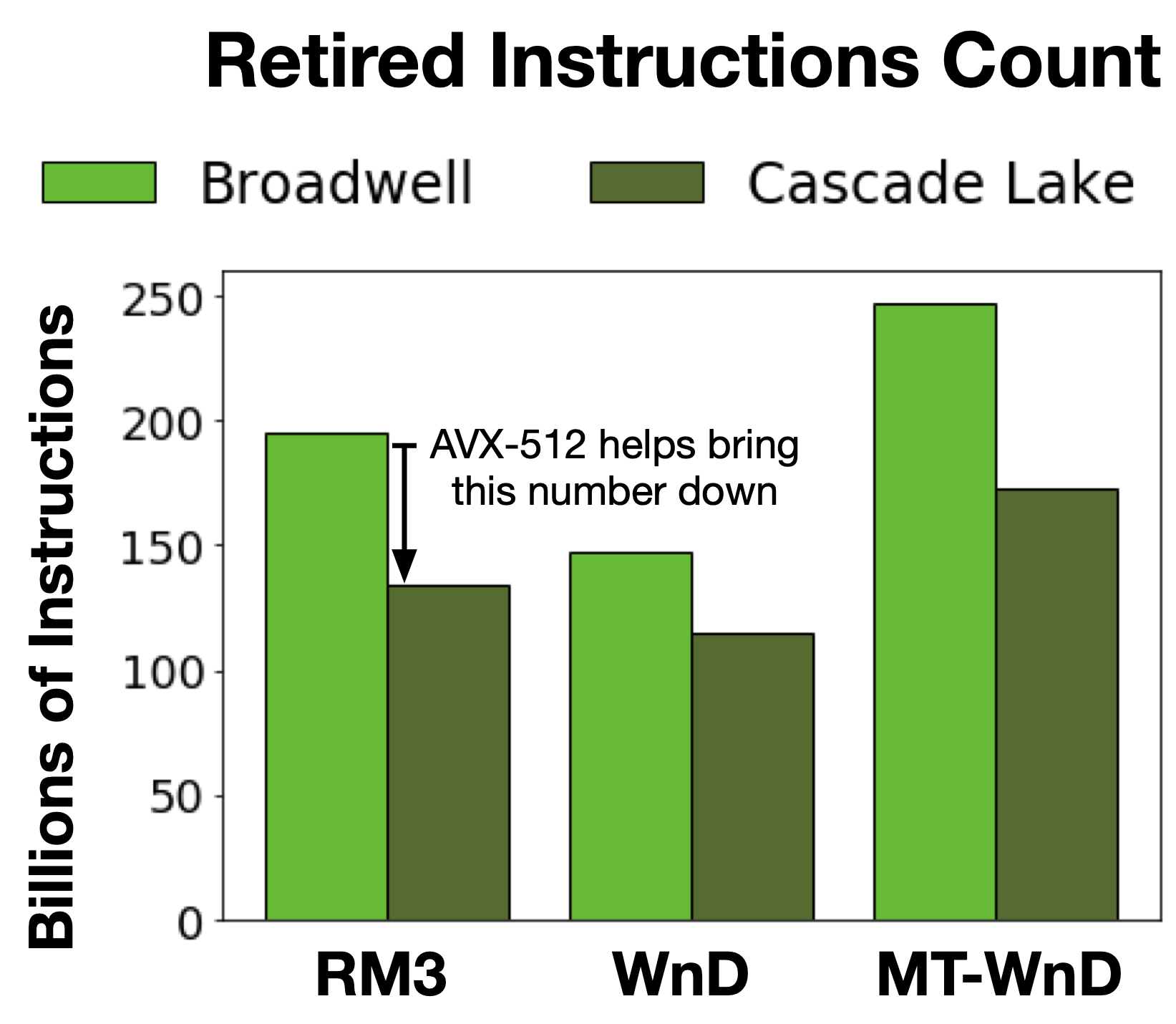}
    \caption{\textbf{Retired Instructions Count} decreases from Broadwell to Cascade Lake due to the introduction of the more efficient AVX-512 VNNI instructions.}
    \label{fig:retired_ins}
    \vspace{-1em}
\end{figure}

\textbf{2) On Cascade Lake, larger FC-dominated models benefit from wider SIMD width and compute capabilities, shifting the bottleneck to the memory subsystem.}

Figure~\ref{fig:td_both} (bottom) shows the TopDown analysis of the eight recommendation models on Cascade Lake CPUs.
In comparison to Broadwell, Cascade Lake enables the majority of models (e.g., NCF, RM1, RM2, DIN, DIEN) to have a larger fraction of retiring pipeline slots.
This larger fraction of cycles spent on retiring instruction is the main reason why Cascade Lake provides consistent speedup over Broadwell (see Figure~\ref{fig:speedups}).
Note that the fraction of cycles devoted to the retiring stage did not increase between Broadwell to Cascade Lake for RM3, WnD, and MT-WnD.
The slight decrease in the retiring cycles is due to fewer total dynamic instructions, as shown in Figure~\ref{fig:retired_ins}; overall, the wider width AVX-512 Vector Neural Network Instructions (VNNI) improves performance for larger FC-dominated models. 

Recall that RM3, WnD, and MT-WnD are core bound on Broadwell.
Figure~\ref{fig:core_bound} (right) shows the backend TopDown analysis on Cascade Lake.
Given the wider AVX512-VNNI instructions, Cascade Lake implements more sophisticated fused multiply-add hardware, which increases the compute capability of the processor.
The increased compute capability reduces pressure on the functional units as shown in Figure~\ref{fig:core_bound} (bottom, right).
Despite the reduced execution port utilization, inference performance for RM3, WnD, and MT-WnD remains backend bound on Cascade Lake.
As shown in Figure~\ref{fig:core_bound} (upper, right), the backend bottleneck has shifted from being core-bound to memory bound.
The particular memory subsystem limiting performance depends on the input batch size -- smaller batch sizes (i.e., less than $100$) are limited by L3 cache accesses while, larger batch-sizes are limited by DRAM latency.

\begin{figure}[t]
    \centering
    \includegraphics[width=0.9\linewidth]{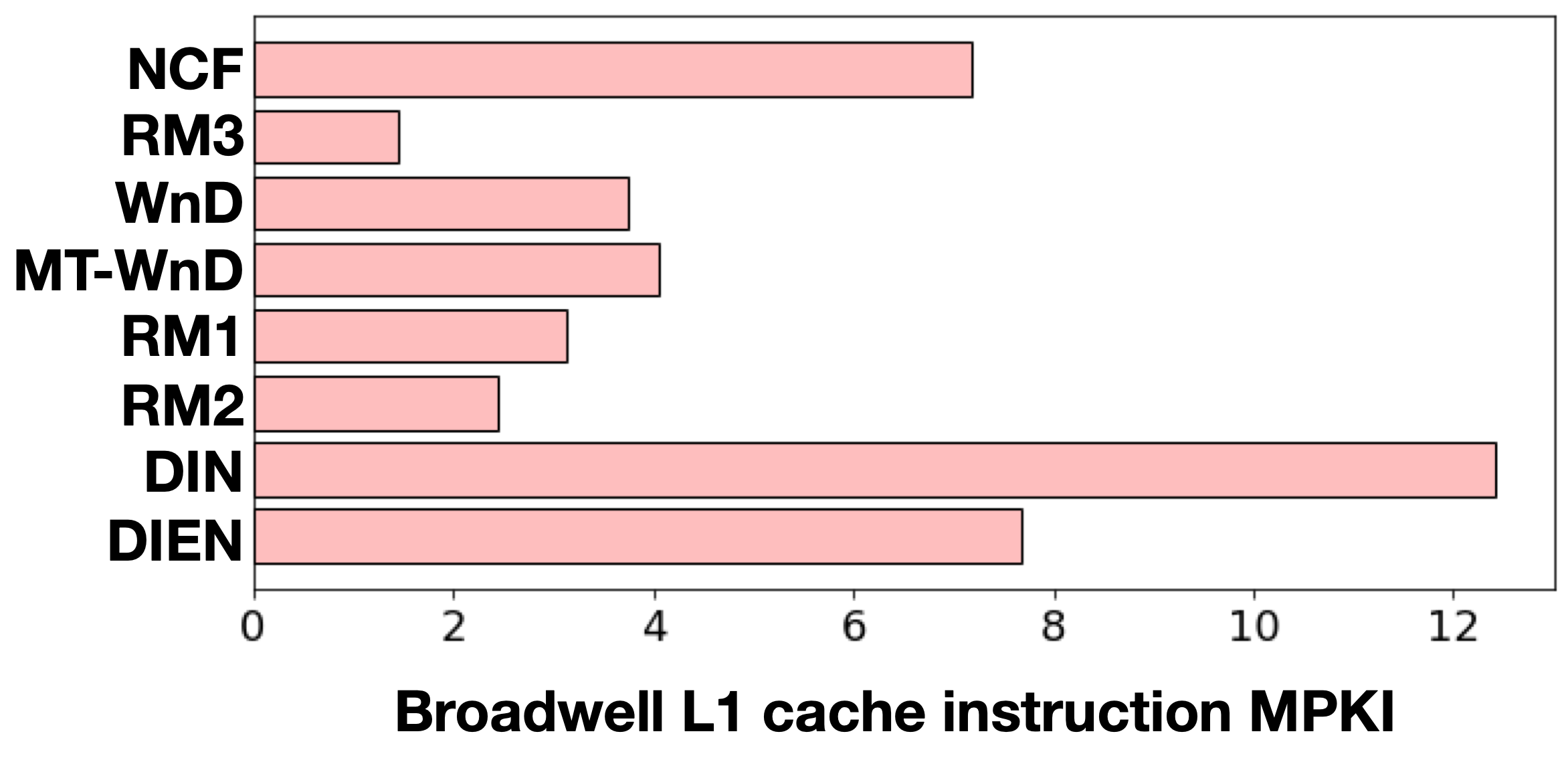}
    \caption{\textbf{NCF and attention-based models suffer from instruction cache misses.} NCF's small size shifts the bottleneck from execution units to i-cache. Attention-based models scale each embedding vector (from irregular memory accesses) with individual weights, leading to high i-cache MPKI.}
    \label{fig:i_mpki_bdw}
    \vspace{-1em}
\end{figure}

\textbf{3) Smaller FC-dominant models and attention-based models suffer from frontend latency -- especially L1 instruction cache latency.} 
Not all FC-dominant models are core-bound. 
For example, on Broadwell, NCF suffers from frontend latency bottlenecks and in particular, L1 instruction cache latency.
Because of its relatively small FC layers, NCF does not exhibit core-bound levels of high compute intensity.

To understand these frontend limitations, Figure \ref{fig:i_mpki_bdw} quantifies the L1 instruction cache miss rate. 
NCF, along with attention-based models like DIN and DIEN, have higher L1 instruction cache miss rates compared to the remaining models.
For instance, we measure a L1 instruction misses per thousand instructions (i-MPKI) of 12.4 and 7.7 for DIN and DIEN, respectively.
The high instruction cache miss rates are tied to how DIN and DIEN implement attention.  
Recall that for recommendation systems, attention allows networks to individually weight the importance of embedding vectors to offer higher personalization. 
In DIN, attention is implemented using hundreds of local concatenation and FC layers; this leads to a large number of instructions with unique reference locations (since the instruction cache does not cache opcodes but specific instructions, including the reference operand).
Given the unique memory addresses for embedding table lookups, the instruction cache hit rate suffers from irregular memory accesses (i.e., lack of spatial and temporal locality).
DIEN's GRU implementation more efficiently translates to matrix operations compared to DIN's implementation with local concatenation-FC \textit{per lookup}. This offers cache friendly loops with regular operand and reference locations.

\begin{figure}[t]
    \centering
    \includegraphics[width=0.9\linewidth]{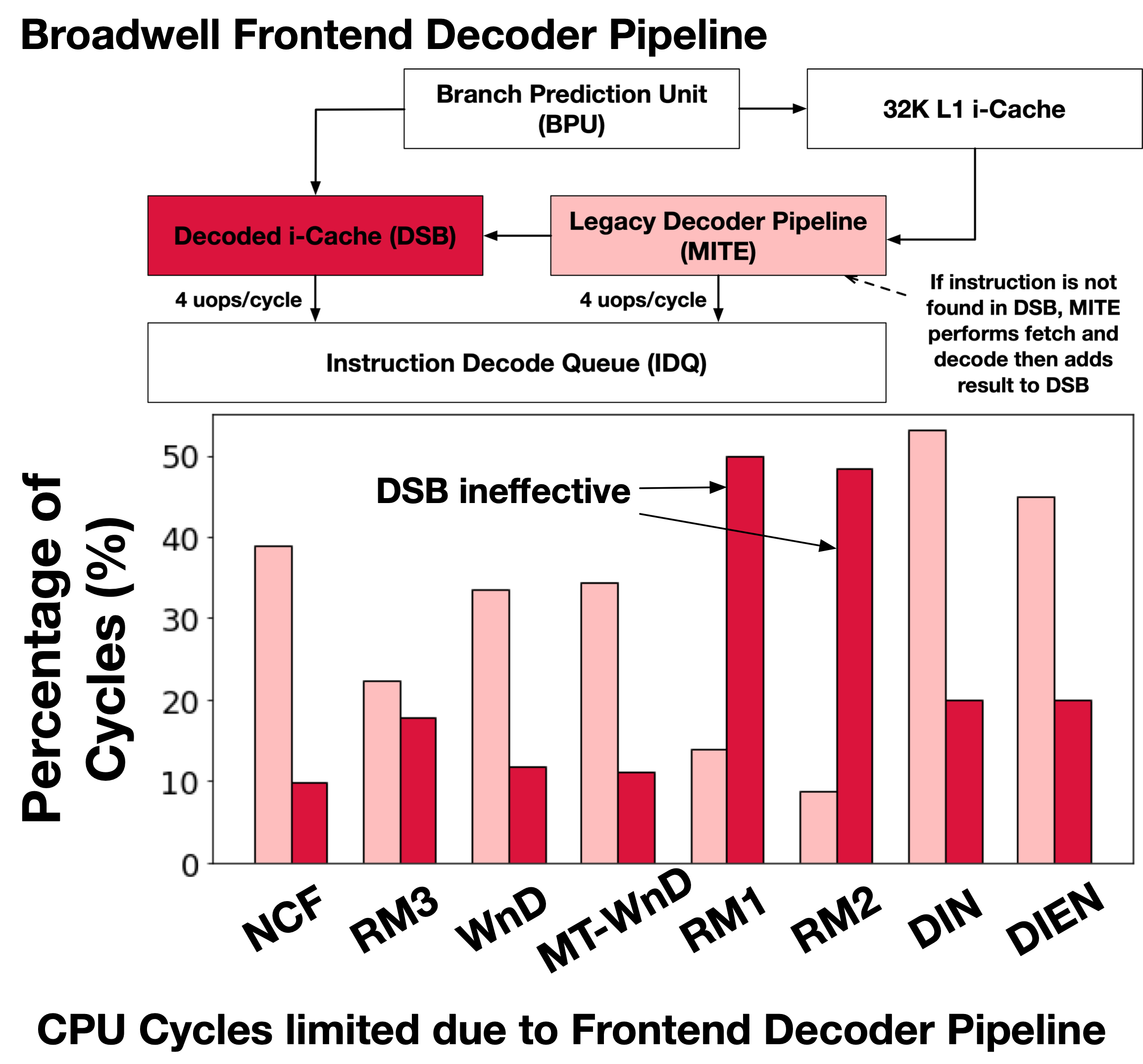}
    \caption{\textbf{Frontend Decoder Pipeline Inefficiencies.} The two main decoder microarchitecture components are DSB and MITE. Shown are percent of cycles in which the CPU was limited by a specific decoder component (i.e., component was not supplying IDQ with optimal number of decoded instructions).}
    \label{fig:decoder_bottleneck}
    \vspace{-1em}
\end{figure}

\begin{figure}[t]
    \centering
    \includegraphics[width=0.4\linewidth]{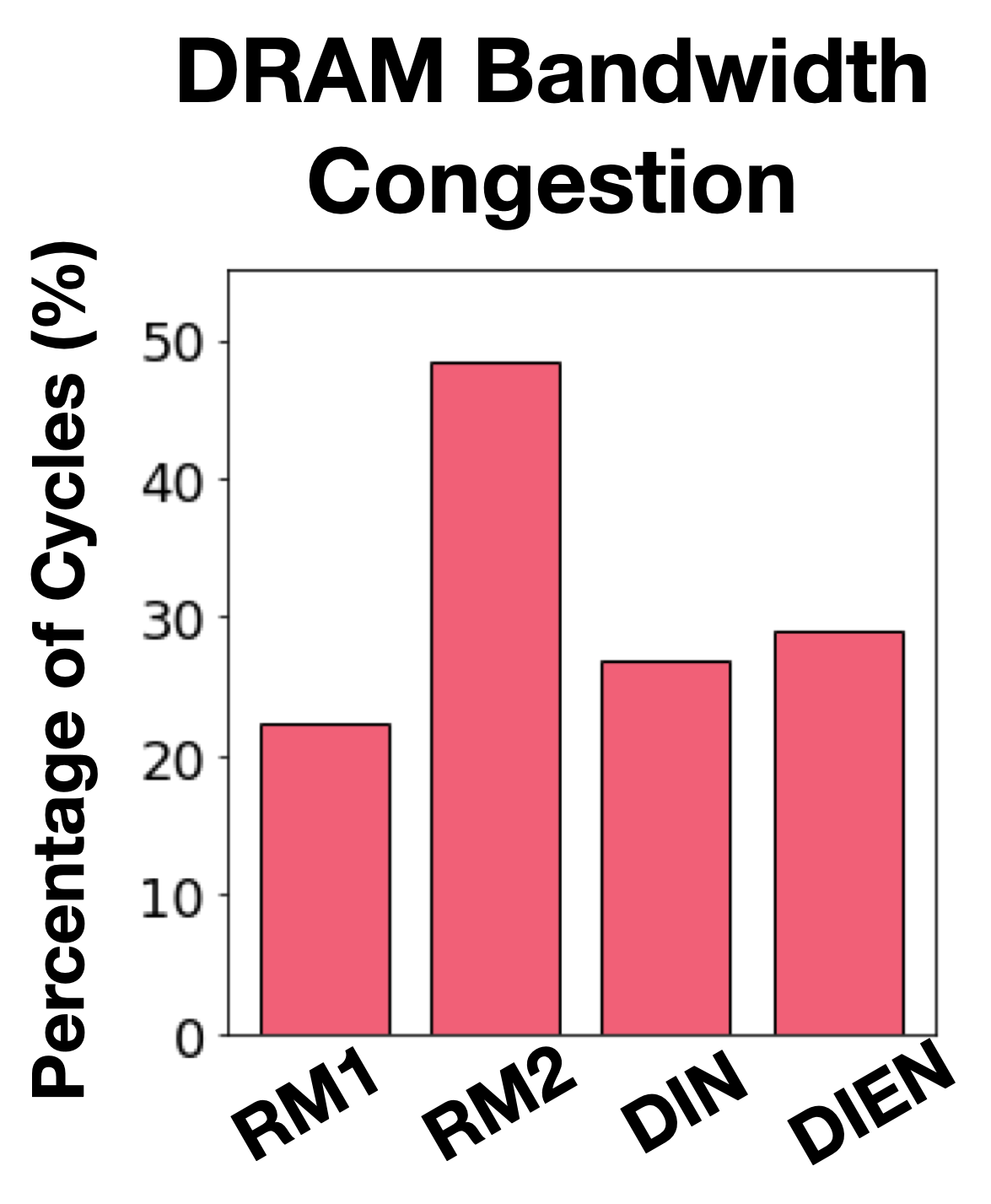}
    \caption{\textbf{RM2 also suffers from DRAM Bandwidth Congestion} from the large number of embedding lookups.}
    \label{fig:dram_bw}
    \vspace{-1em}
\end{figure}

\textbf{4) Models with more embedding table lookups suffer from instruction decoder bottlenecks. As the degree of embedding table lookups increases, the performance bottleneck shifts from pure decoder issues to also include DRAM bandwidth limitations.}
On Broadwell, RM1 and RM2 are frontend bandwidth-bound deep recommendation models. 
Generally, this denotes inefficiencies in the instruction decode phase as opposed to in the instruction fetch phase.
Figure \ref{fig:decoder_bottleneck} illustrates the fraction of cycles spent on two parts of Broadwell's frontend decoder pipeline, the decoded i-cache (Decoded Stream Buffer - DSB) and the legacy decoder pipeline (Micro-Instruction Translation Engine - MITE). 
MITE is responsible for fetching instructions from instruction memory and decoding them into $\mu$ops while the DSB caches results from MITE.
For each target instruction, DSB is first queried.
If the instruction is found in the DSB, the corresponding $\mu$ops are directly delivered to the instruction decode queue (IDQ).
If the instruction is not found, MITE is used to fetch and decode instructions and the result is added into DSB.

Figure \ref{fig:decoder_bottleneck} (bottom) shows Broadwell's decoder pipeline -- including DSB and MITE. 
CPU cycles are analyzed to determine if either DSB or MITE could not supply IDQ with sufficient $\mu$ops.
For both RM1 and RM2, the frontend bandwidth bound models, TopDown analysis illustrates the bottlenecks in DSB as the main source of inefficiency.

The DSB bottleneck can be tied to algorithmic, model-architecture features of RM1 and RM2.
In particular, both models require a high degree, tens to hundreds, of embedding table lookups.
Combined with the irregular memory accesses coming from embedding lookups, larger instruction footprints stress the DSB.
Furthermore, RM1 and RM2 spend a large fraction of their cycles on bad speculation as shown in Figure~\ref{fig:td_both}.
Since DSB is also affected by the Branch Prediction Unit (BPU), the large amount of branch misprediction latency will degrade the performance of DSB, as the speculation stalls are primarily from branch mispredictions.

Despite the similarities between RM1 and RM2, RM2 has a unique performance bottleneck.
As shown in Table~\ref{tab:rec_models}, RM2 comprises more embedding tables (32 versus 8 in RM1) and more lookups per table (120 versus 80 in RM1). 
Given the larger size, RM2 suffers from bottlenecks in both the frontend and backend pipeline.
Figure~\ref{fig:dram_bw} illustrates the DRAM bandwidth congestion of RM1, RM2, DIN, and DIEN.
DRAM bandwidth congestion, as defined by Intel, occurs when the offcore read queue occupancy exceeds 70\% of the maximum number of requests that can be served by the memory controller simultaneously; whereas when below 70\% occupancy, the stall can be characterized as DRAM latency bound~\cite{intel2020opt}.
We find that RM2 suffers from significantly higher DRAM bandwidth congestion limitations compared to the other models.
Previous work exploit this property to design near memory processing solutions for DRAM bandwidth-bound recommendation models~\cite{rhu2019tensordimm, ke2019recnmp}.

\begin{figure}[t]
    \centering
    \includegraphics[width=\linewidth]{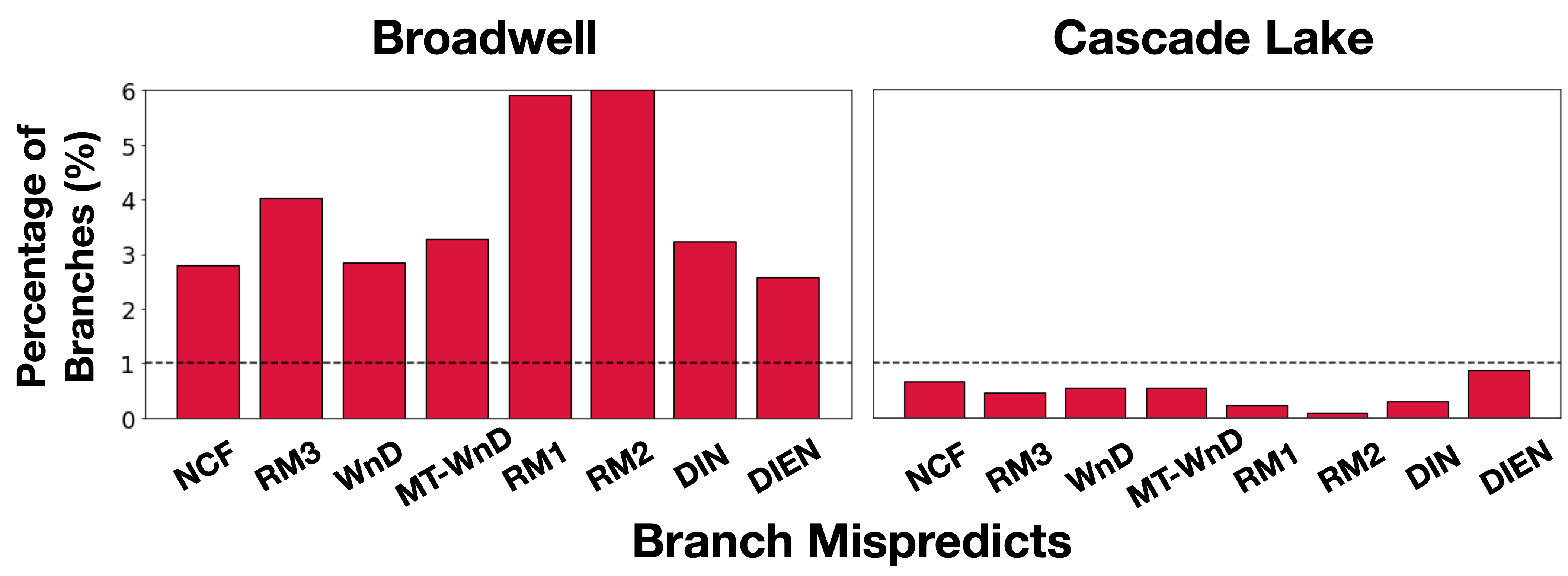}
    \caption{\textbf{Branch Mispredicts} decrease significantly when we transition from Broadwell to Cascade Lake machines.}
    \label{fig:branches}
    \vspace{-1em}
\end{figure}

\textbf{5) Cascade Lake significantly reduces the amount of pipeline slots lost to bad speculation.} 
One of the marked differences between the TopDown breakdowns of Broadwell and Cascade Lake in Figure \ref{fig:td_both} is the decrease in pipeline slots lost to bad speculation in Cascade Lake. 
While the specific detail of the branch predictor designs used in Broadwell and Cascade Lake are not available, the transition from Broadwell to Skylake sees a penalty reduction for incorrect direct jump target~\cite{fog2020microarch}. 
This overall improvement shifts the Cascade Lake backend bottlenecks to the memory subsystems as discussed in Observation \#2.

\subsection{Tying Model Architectures to Pipeline Bottlenecks}

To tie our microarchitectural observations to the specifics of recommendation model architectures, we quantify the effects of select algorithmic model architecture features with a linear regression model.

\begin{figure}[t]
    \centering
    \includegraphics[width=\linewidth]{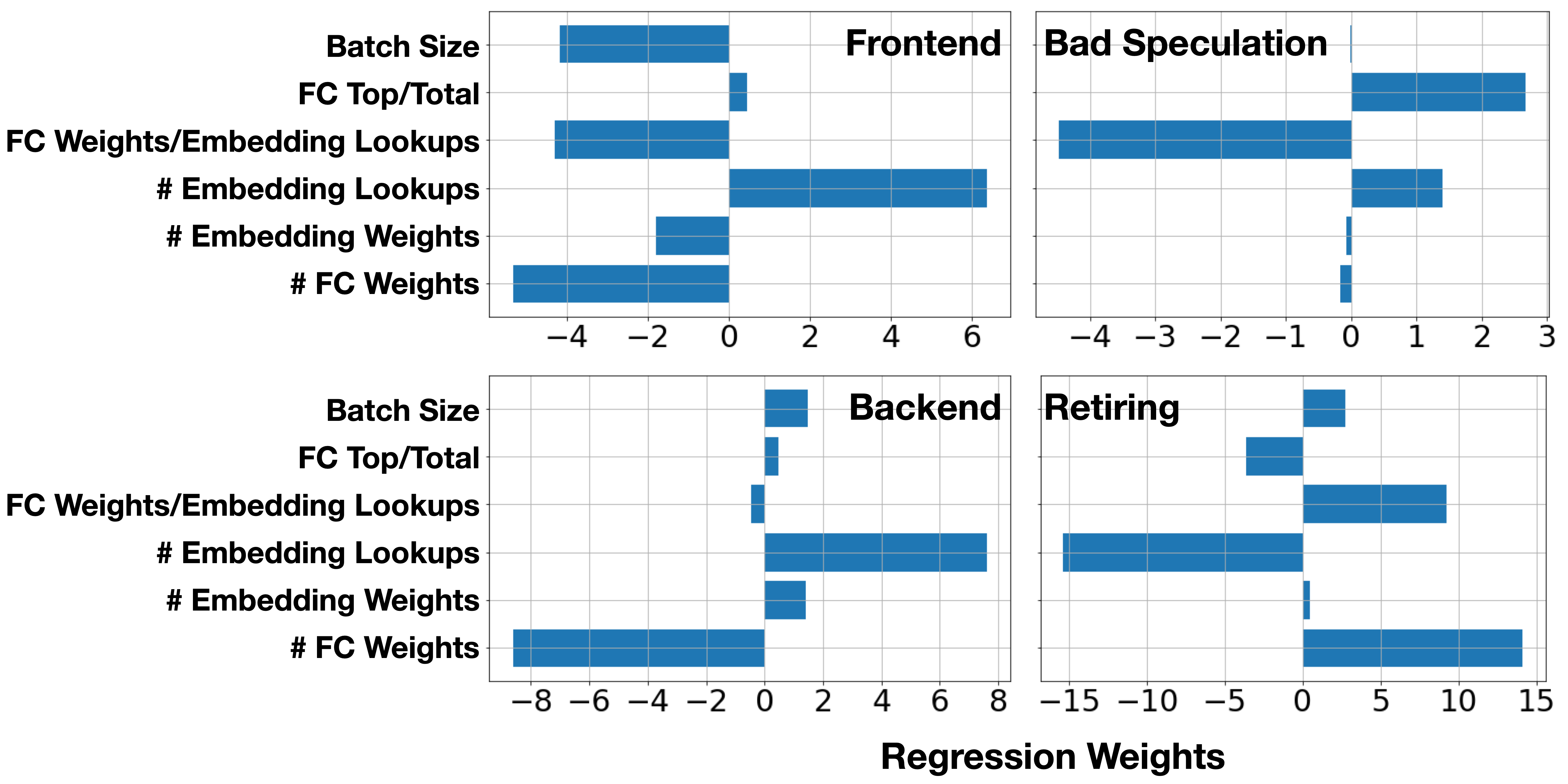}
    \caption{\textbf{Linear Regression modeling} of algorithmic model architecture components and pipeline bottlenecks reveals that there is not a single deciding factor for each bottleneck.}
    \label{fig:lr_weights}
\end{figure}

Figure \ref{fig:lr_weights} summarizes our linear regression modeling; all input features have been normalized so the weight magnitude represents degree of impact.
Data points are collected from running the 8 models at batch sizes from 1 to 16384. The model shows that each pipeline bottleneck is a result of a combination of different algorithmic model architecture features. For example, this model shows that a high ratio of FC to embedding weights reduces bad speculation while a top-heavy distribution of FC weights leads to increases in bad speculation. The first point explains the intuition that compute-intensive models have more predictable branches while the second point shows that more direct processing of continuous inputs is correlated with less bad speculation.




%% file: text/discussion.tex




%% file: text/related_work.tex
\section{Related Work}~\label{sec:related_work}
\textbf{Analysis and optimizations for recommendation systems.} Recommendation systems have recently come under the spotlight for computer systems researchers. As mentioned earlier, a few recent works explore near-memory processing techniques for recommendation models dominated by table lookup operations. 
TensorDimm evaluates near-data processing enabled custom DIMM modules on recommendation models similar to RM1-3~\cite{rhu2019tensordimm}; 
RecNMP evaluates a set of techniques centered around memory-side caching on production-representative embedding traces~\cite{ke2019recnmp}.
Ginart et al. and Shi et al.~\cite{ginart2019mixed, shi2019compositional} compresses embedding tables in recommendation models while maintaining the model accuracy.
Centaur extends near-memory processing designs to also account for the MLP layers through a chiplet-based accelerator design~\cite{hwang2020centaur}.
Other works have explored at-scale optimizations~\cite{gupta2020architectural}:
DeepRecSys explores different optimizations at the datacenter scale; the recommendation suite evaluated throughout this paper is from DeepRecSys's open sourced implementations~\cite{gupta2020deeprecsys}. Other work has started to explore implications of training~\cite{kalamkar2020optimizing, naumov2020fbtraining}.
In contrast, this paper focuses on purely characterizing the recommendation suite introduced in~\cite{gupta2020deeprecsys}. To the best of our knowledge, this is the first detailed microarchitectural characterization of deep recommendation systems.

\textbf{DNN benchmarks and accelerator designs.} Current benchmarks and characterizations for DNNs primarily focus on FC, CNNs, and RNNs~\cite{adolf2016fathom,wei2019benchmarking,coleman2017dawnbench,zhu2018benchmarking, mlperf-reco-advisory, mlperf}. Building upon the performance bottlenecks derived from these studies, a variety of hardware solutions have been proposed to optimize for traditional DNNs~\cite{scnn,rhu2016vdnn,kwon2018beyond,choi2019prema,gupta2019masr,eyeriss,eie, minerva,gonzalez2018pact,fletcher2018micro,fletcher2019micro,dadiannao,cambricon,sharma2016dnnweaver,du2015shidiannao,liu2015pudiannao,chi2016prime,shafiee2016isaac,kim2016neurocube,likamwa2016redeye,mahajan2016tabla,gao2017tetris,maxnvm,albericio2017bit}. 
While these DNNs share the operators introduced in Section~\ref{sec:software}, recommendation models present them in unique ratios and model architecture organizations.

%% file: text/conclusion.tex
\section{Conclusion}~\label{sec:conclusion}
It is important to characterize deep recommendation models across different layers in the execution stack because this helps us better understand the bottlenecks that arise from our evaluations. By understanding more about these bottlenecks and how they realize themselves at different levels (i.e., as operators in Caffe2 and as inefficiencies of different CPU components), we can intelligently design future hardware that optimizes for deep recommendation inference.

%% file: text/acknowledgements.tex
\section{Acknowledgements}~\label{sec:acknowledgements}
We would like to thank the anonymous reviewers for their thoughtful comments and suggestions. 
We would also like to thank Glenn Holloway and Emma Wang for their valuable feedback.
This work was sponsored in part by NSF CCF-1533737 and a National Science Foundation Graduate Research Fellowship (NSFGRFP).